\documentclass[submission, Phys]{SciPost}

\pdfoutput=1
\usepackage{comment}
\usepackage{enumerate}
\usepackage{amssymb}
\usepackage{amsmath}
\usepackage{braket}
\usepackage{graphicx}
\usepackage{dsfont}
\usepackage{booktabs}

\newcommand{\sech}{\mathrm{sech}}
\newcommand{\csech}{\mathrm{cosech}}

\newcommand{\be}{\begin{equation}}
\newcommand{\ee}{\end{equation}}
\newcommand{\ba}{\begin{aligned}}
\newcommand{\ea}{\end{aligned}}
\newcommand{\bw}{\begin{widetext}}
\newcommand{\ew}{\end{widetext}}

\newcommand{\bea}{\begin{eqnarray}}
\newcommand{\eea}{\end{eqnarray}}

\begin{document}

\begin{center}{\Large \textbf{
Free fermions with dephasing and boundary driving: Bethe Ansatz   results
}}\end{center}

\begin{center}
Vincenzo Alba\textsuperscript{1*}
\end{center}

\begin{center}
{\bf 1}	
Dipartimento di Fisica dell' Universit\`a di Pisa and INFN, Sezione di Pisa, I-56127 Pisa, Italy\\
* vincenzo.alba@unipi.it
\end{center}

\begin{center}
\today
\end{center}


\section*{Abstract}
{\bf By employing the Lindblad equation, we derive the evolution of the two-point correlator for a free-fermion 
	chain of length $L$ subject to  bulk dephasing and 
	boundary losses. 
	We use the Bethe ansatz to diagonalize the Liouvillian ${\mathcal L}^{\scriptscriptstyle(2)}$  
	governing the dynamics of the correlator.
	The majority of its  eigenvalues  are complex. 
	Precisely, $L(L-1)/2$ complex eigenvalues do not depend on dephasing, 
	apart from a trivial shift. The remaining complex levels 
	are perturbatively related to the dephasing-independent ones for large $L$.  
	The long-time dynamics is governed by a band of real eigenvalues, which contains 
	an extensive number of levels.  
	They give rise to diffusive scaling at intermediate times, when boundaries can be 
	neglected. Moreover, they encode the breaking of diffusion at asymptotically long times. 
	Interestingly, for large loss rate two boundary modes appear in the spectrum. 
	The real eigenvalues correspond to string solutions of the Bethe equations, and  
	can be treated effectively for large chains. This allows us to derive compact formulas 
	for the dynamics of the fermionic density. We check our results against exact diagonalization, 
	finding perfect agreement. 
}

\section{Introduction}
\label{sec:model}

Markovian master equations~\cite{petruccione2002the}, such as the Lindblad equation, provide a versatile 
tool to understand the interplay between coherent and dissipative dynamics in \emph{open} quantum 
many-body systems~\cite{rossini2021coherent}.  
Although the interaction with an environment typically is a strong adversary for quantum coherence, it 
can also be exploited to imprint nontrivial quantum correlations~\cite{lin2013}, to aid  
quantum computation~\cite{verstraete-2009}, or to stabilize topological order~\cite{diehl-2011}. 

Exact solvable models could potentially help to build a general understanding 
of open quantum systems, similar to what happened in out-of-equilibrium \emph{closed} systems~\cite{calabrese2016introduction}. 
Unfortunately, despite intense effort there are comparatively  
few examples of exact solvable Lindblad equations. Free-fermion and free-boson 
models subject to arbitrary \emph{linear} jump operators lead to quadratic Liouvillians, and 
stand out as prominent examples~\cite{prosen2008third}. 
Still, non quadratic Liouvillians that are solvable, for instance by the Bethe ansatz, 
exist~\cite{znidaric-2010,znidaric-2011,prosen2011exact,medvedyeva2016exact,ilievski2017dissipation,buca2020bethe,essler2020integrability,ziolkowska2020yang,robertson2021exact,nakagawa2021exact,yamamoto2022universal,shibata2020quantum,deleeuw2021contructing,claeys2022dissipative,deleeuw2022bethe,yoshida2023liouvillian,yamanaka2023exact}. 
Remarkably, it has been shown in Ref.~\cite{medvedyeva2016exact} that the Liouvillian describing the 
out-of-equilibrium dynamics of the fermionic tight-binding chain with global dephasing can 
be mapped to the Hubbard chain with imaginary interaction strength, which  can be solved by Bethe ansatz~\cite{essler2005the}. 
Interestingly, it is well-established that the dynamics of simple observables, such as few-point correlation 
functions can be obtained analytically~\cite{esposito2005emergence,esposito2005exactly,eisler2011crossover}, 
without explicitly relying on the exact solvability of the full Liouvillian. 
Furthermore, integrability is crucial to devise effective descriptions for 
out-of-equilibrium open systems. For instance, it has been shown recently 
that, by exploiting integrability, the Lindblad dynamics of 
paradigmatic observables can be  captured within the hydrodynamic 
framework~\cite{bouchoule2020the,rossini2021strong,riggio2023effects,coppola2023wigner,perfetto2023reaction,perfetto2023quantum}. 
For quadratic Liouvillians one can employ the 
quasiparticle picture~\cite{calabrese2005evolution,fagotti2008evolution,alba2017entanglement} 
to describe the dynamics of entanglement-related 
quantities~\cite{alba2021spreading,carollo2022dissipative,alba2022hydrodynamics,alba2022logarithmic}. 
Similar results were derived for free fermions in the presence of 
localized dissipation~\cite{alba2022noninteracting,alba2021unbounded,caceffo2023entanglement}. 
In conclusion, widening the set of integrable Lindblad equations is of 
paramount importance to make progress in out-of-equilibrium open quantum systems. 
%
\begin{figure}[t]
\centering
\includegraphics[width=.7\linewidth]{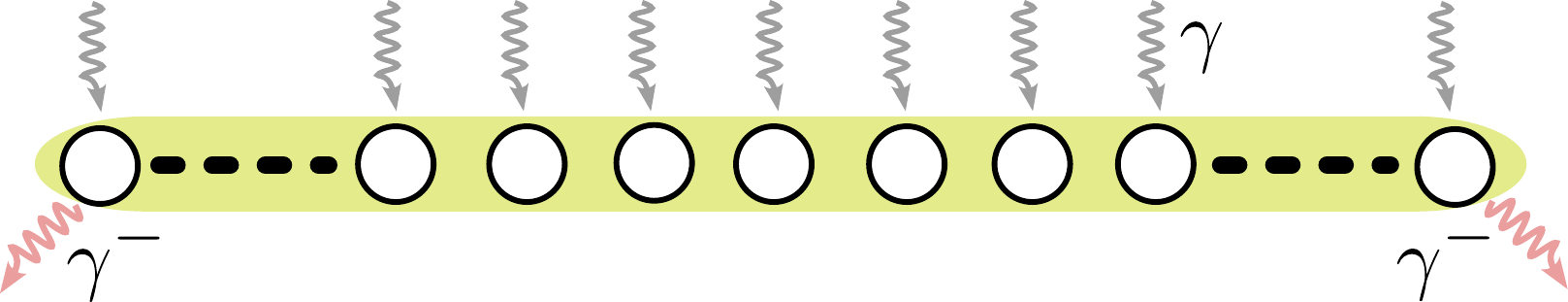}
\caption{ Setup considered in this paper. A tight-binding fermionic chain 
 with $L$ sites is subject to global dephasing, with dephasing rate $\gamma$. We employ 
 open boundary conditions. At the edges of the chain fermions are incoherently removed at 
 rate $\gamma^-$. 
}
\label{fig0:cartoon}
\end{figure} 
%

Here we provide new results in this direction considering 
the setup depicted in Fig.~\ref{fig0:cartoon}. We focus on a fermionic tight-binding chain 
of length $L$ with open boundary conditions. Global dephasing is present on each site of 
the chain. We denote by $\gamma$ the dephasing rate. Besides, the 
chain is subject to incoherent fermion losses with the same rate 
$\gamma^-$ at the edges. 
This prototypical setup was investigated in Ref.~\cite{znidaric-2010}, and more 
recently in Ref.~\cite{turkeshi2021diffusion,turkeshi2022enhanced} 
(see also Ref.~\cite{jin-2020,jin2022exact}) and~\cite{tang2024quantum,kattel2024spin}. The focus was on 
the interplay between the diffusive transport induced by the global 
dephasing and the ballistic one due to the boundary driving. 
The same setup was employed 
to study the interplay between dissipation and criticality~\cite{rossini2021coherent,tarantelli2021quantum}. 
Here we focus on the two-point fermionic correlation function $G_{x_1,x_2}:=\mathrm{Tr}(\rho(t)c^\dagger_{x_1} c_{x_2})$, 
with $c_{x_1},c_{x_2}$ standard fermion operators, and $\rho(t)$ the full-system density matrix. 
The out-of-equilibrium dynamics of $G_{x_1,x_2}$ is governed by a Liouvillian super operator ${\mathcal L}^{\scriptscriptstyle(2)}$ 
(see section~\ref{sec:model}). 

We show that the spectrum, i.e., eigenvalues and eigenvectors, of ${\mathcal L}^{\scriptscriptstyle(2)}$  can be constructed 
explicitly using the Bethe ansatz. This happens despite the fact that the full Liouvillian, 
to the best of our knowledge, is not integrable. 
Indeed, as we show, the full Liouvillian is mapped to the one-dimensional 
Hubbard model with imaginary density-density interaction, imaginary boundary magnetic fields,  
and imaginary  boundary pair production. The last term creates a pair of fermions with opposite spins 
at the boundaries, making the model non integrable. 
Since ${\mathcal L}^{\scriptscriptstyle (2)}$ is not hermitian, its eigenvalues (energies) are in general 
complex. The spectrum of ${\mathcal L}^{\scriptscriptstyle(2)}$, at least for moderate dephasing rate, 
splits into three different components. 
We show that there are $L(L-1)/2$ complex eigenvalues that are trivially related by a shift by 
$-\gamma$ to the eigenvalues obtained in the absence of dephasing. 
These correspond to eigenstates that are free-fermionic in nature. 
For $\gamma^-=0$ these eigenvalues 
form a vertical band in the complex plane, their real part being $-\gamma$. For nonzero 
$\gamma^-$ the band is deformed. Nearby in energy, there are $\sim L(L-1)/2$ 
complex eigenvalues, which become the same as  the dephasing-independent ones 
in the large $L$ limit. 

Finally, a band containing $\sim L$ real eigenvalues is present. 
A similar band is present for periodic boundary conditions (see Ref.~\cite{esposito2005emergence} and Ref.~\cite{eisler2011crossover}), 
where it is responsible for diffusive dynamics at long times. For this reason we dub it diffusive band.  
The eigenvalues with the largest real parts, and the gap of ${\mathcal L}^{\scriptscriptstyle(2)}$, are in the diffusive band. 
The diffusive band correspond to so-called string solutions of the Bethe equations,  
which in the large $L$ limit can be treated within the framework of the string hypothesis~\cite{essler2005the}. 
For instance, this allows us to derive the  Liouvillian gap analytically  
as $\Delta {\mathcal L}^{\scriptscriptstyle(2)}=-2\pi^2/(\gamma L^2)+\beta/L^3+{\mathcal O}(L^{-4})$, 
where the constant $\beta$, which we determine, depends both on $\gamma$ and $\gamma^-$. 
The  number of eigenvalues in the diffusive band depends on $\gamma,\gamma^-$. We show that in the large 
$L$ limit, i.e., in the regime of validity of the string hypothesis, 
there is a ``critical'' $\gamma_c=4$ above which the band contains the largest number of eigenvalues. 
Upon lowering $\gamma$ the band gets progressively depleted. In the limit $L\to\infty$ the energy 
$\varepsilon$ in the diffusive band are such that $\varepsilon>\sqrt{\gamma^2-16}-\gamma$ for 
$\gamma>\gamma_c$. The number of levels depends on $\gamma^-$ as well. Precisely, for $\gamma>\gamma_c$ the band 
contains $L$ levels for $\gamma^-<\gamma^-_c=\exp(-\mathrm{arccosh}(\gamma/4))$. At larger $\gamma^-$, 
two of the eigenvalues detach from the diffusive band, and are pushed to lower 
$\mathrm{Re}(\varepsilon)$ upon increasing 
$\gamma^-$. The splitting between them is exponentially suppressed with $L$. These 
eigenvalues  correspond to edge modes of ${\mathcal L}^{\scriptscriptstyle(2)}$, and 
are reminiscent of the boundary-related eigenstates of the Hubbard chain with boundary magnetic 
fields~\cite{essler2005the}.

The Bethe ansatz diagonalization of ${\mathcal L}^{\scriptscriptstyle(2)}$ allows, in principle, 
to obtain the full-time dynamics of  $G_{x_1,x_2}(t)$. This is not straightforward 
because it requires to extract all the $L^2$ solutions of the Bethe equations. Still, in the long-time 
limit the dynamics of the correlator is determined by the diffusive band, which can be treated by using the 
string hypothesis. Here we provide compact expressions for the dynamics of the density profile starting from a fermion 
localized at an arbitrary site of the chain. This is the main ingredient to obtain the dynamics from 
an arbitrary initial density profile. At long times, but short enough that the boundaries can be neglected, 
the density profile exhibits the same diffusive scaling as for periodic boundary conditions. At long 
time the diffusive regime breaks down due to the boundary losses. 

The manuscript is organized as follows. In section~\ref{sec:model} we introduce the tight-binding 
chain with dephasing and boundary losses, and the Lindblad equation. In section~\ref{sec:ba} 
we present the Bethe ansatz treatment of the Liouvillian ${\mathcal L}^{\scriptscriptstyle(2)}$. 
Specifically, in section~\ref{sec:bethe-states} we introduce the ansatz for the eigenstates of 
${\mathcal L}^{\scriptscriptstyle(2)}$. In section~\ref{sec:bethe-equation} we derive the Bethe 
equations. In section~\ref{sec:deph-ind} we discuss the eigenvalues that correspond to dephasing-independent 
solutions of the Bethe equations. 
In section~\ref{sec:vanishing} we investigate the eigenvalues of ${\mathcal L}^{\scriptscriptstyle(2)}$ 
that are perturbatively connected to the dephasing-independent ones 
in the large $L$ limit. Finally in section~\ref{sec:diff} we discuss the diffusive band. 
In section~\ref{sec:asy} we focus on the dynamics of the two-point fermionic 
correlation function. In particular, in section~\ref{sec:left} we discuss how to expand the 
initial correlator in the basis of  the Bethe states. 
In section~\ref{sec:norm} we derive the normalization of the Bethe states. 
In section~\ref{sec:green}, by using the string hypothesis, we derive the long-time 
limit  of the fermion density profile. We discuss numerical results in section~\ref{sec:numerics}. 
Precisely, in section~\ref{sec:beq-num} we focus on the solution of the Bethe equations. We 
provide the full set of solutions for chains with $L=2$ and $L=3$. In section~\ref{sec:bgt-eq} 
we discuss the solution of the Bethe-Gaudin-Takahashi equation for the  diffusive band. In section~\ref{sec:liouv} 
we overview the general structure of the eigenvalues of ${\mathcal L}^{\scriptscriptstyle(2)}$ presenting 
exact diagonalization (ED) data. 
In section~\ref{sec:ed-vs-bethe} we compare the ED data against Bethe ansatz results. 
In section~\ref{sec:gap} we focus on the finite-size scaling of the 
Liouvillian gap. In section~\ref{sec:num-G} we benchmark the Bethe ansatz results for 
$G_{x_1,x_2}$ with exact diagonalization. Section~\ref{sec:profile-den} provides 
numerical results for the dynamics of the density profile. In section~\ref{sec:scaling} we 
focus on the diffusive scaling of the fermion density and its violation due to the boundary losses. 
We conclude in section~\ref{sec:conc}. In Appendix~\ref{app:mapping} we show that the full Liouvillian 
of the system is mapped to a one-dimensional Hubbard model with imaginary interaction, imaginary boundary 
fields and imaginary boundary fermion pair production. In Appendix~\ref{app:hubb} we compare the Bethe equations 
for the Hubbard chain with boundary fields and the Bethe equations derived in section~\ref{sec:bethe-equation}, 
showing that they are equivalent.

\section{Free fermions with dephasing and boundary losses}
\label{sec:model}

Here we consider the fermionic tight-binding chain described by the Hamiltonian 
\begin{equation}
	\label{eq:ham}
	H=\sum_{x=1}^{L-1} (c^\dagger_xc_{x+1}+c^\dagger_{x+1}c_x),  
\end{equation}
where $c^\dagger_x,c_x$ are standard fermionic creation and annihilation operators. 
The system lives on a chain with  $L$ sites. We employ open boundary conditions. 
Our setup is depicted in Fig.~\ref{fig0:cartoon}. 
The chain undergoes a nonunitary dynamics described by the Lindblad master equation~\cite{petruccione2002the}. 
The state of the system is described by a density matrix $\rho(t)$. Within the 
framework of Markovian master equations~\cite{petruccione2002the}, the dynamics of $\rho$  
is obtained by solving the Lindblad equation as 
\begin{equation}
	\label{eq:lind}
	\frac{d\rho(t)}{dt}:= {\mathcal L}(\rho)=-i[H,\rho(t)]+\sum_{x=1}^L \sum_\alpha\Big(L_{x,\alpha}\rho(t) L_{x,\alpha}^\dagger- 
	\frac{1}{2}\big\{L_{x,\alpha}^\dagger L_{x,\alpha},\rho(t)\big\}\Big), 
\end{equation}
where $\{,\}$ denotes the anticommutator, and $L_{x,\alpha}$ is the so-called Lindblad operator acting at site 
$x$. In~\eqref{eq:lind}, ${\mathcal L}$ is the Liouvillian. 
The label $\alpha$  encodes the different types of dissipation. Specifically, we choose $\alpha=1$ for 
global dephasing and $\alpha=2$ for boundary losses. 
The Lindblad operator for global dephasing reads as 
\begin{equation}
	\label{eq:l-deph}
	L_{x,1}=\sqrt{\gamma} c^\dagger_x c_x, \quad x\in[1,L]. 
\end{equation}
Localized losses at the edges of the chain are described by 
\begin{equation}
	\label{eq:l-loss}
	L_{x,2}=\sqrt{\gamma^-} c_x, \quad x=1,L, 
\end{equation}
In~\eqref{eq:l-deph} and~\eqref{eq:l-loss}, $\gamma$ and $\gamma^-$ are the 
dephasing and loss rates, respectively. By using~\eqref{eq:lind}, it is straightforward to obtain the evolution of the 
fermionic two-point correlation function $G_{x_1,x_2}(t)$ defined as~\cite{landi2022nonequilibrium}  
\begin{equation}
	\label{eq:gxy}
	G_{x_1,x_2}(t):=\mathrm{Tr}(\rho(t) c^\dagger_{x_1}c_{x_2}). 
\end{equation}
The dynamics of $G_{x_1,x_2}$ from a generic initial condition $G_{x_1,x_2}(0)$ is 
obtained by solving the system of equations as~\cite{alba2021spreading}  
\begin{multline}
	\label{eq:G}
	\frac{dG_{x_1,x_2}}{dt}:={\mathcal L}^{(2)}(G_{x_1,x_2})= i(G_{x_1-1,x_2}+G_{x_1+1,x_2}-G_{x_1,x_2-1}-G_{x_1,x_2+1}) 
	\\-\gamma G_{x_1,x_2}(1-\delta_{x_1,x_2}) 
	-\gamma^- G_{x_1,x_2}(\delta_{x_1,1}+\delta_{x_2,1}+\delta_{x_1,L}+\delta_{x_2,L}), 
\end{multline}
where we define the $L^2\times L^2$ linear super operator ${\mathcal L}^{\scriptscriptstyle(2)}$. 
The superscript in ${\mathcal L}^{\scriptscriptstyle(2)}$ is to stress that 
${\mathcal L}^{\scriptscriptstyle(2)}$ is not the same Liouvillian appearing in~\eqref{eq:lind}, which is a 
$2^L\times 2^L$ matrix and governs the dynamics of the full-system density matrix. 
In the following we will refer to ${\mathcal L}^{\scriptscriptstyle(2)}$ as the 
Liouvillian, and to ${\mathcal L}$ (cf.~\eqref{eq:lind}) as the full Liouvillian. 
In~\eqref{eq:G} we consider the symmetric situation in which the fermions 
are removed at the edges of the chain at the same rate $\gamma^-$. However, the case with 
different rates $\gamma^-_{\scriptscriptstyle L(R)}$ at the left and right edges 
of the chain can be considered as well. 
The first term in~\eqref{eq:G} describes the unitary dynamics governed by the 
Hamiltonian~\eqref{eq:ham}. The second term is the dephasing, which suppresses the off-diagonal 
elements of $G_{x_1,x_2}$. The last term describes incoherent absorption of fermions at the 
edges of the chain.

Importantly, the solution of~\eqref{eq:G} allows one to obtain the dynamics of $G_{x_1,x_2}$ in 
several physical situations. For instance, let us consider the setup investigated in 
Ref.~\cite{turkeshi2021diffusion}, in which a free-fermion chain is  
subject to global dephasing and fermion pumping at the left edge of the chain and fermion 
losses at the right one. Let us focus on the case with dephasing rate $\gamma$, and 
equal pump/loss rate $\gamma'$. The evolution of $G_{x_1,x_2}$ is obtained by 
solving~\cite{turkeshi2022enhanced} 
\begin{multline}
	\label{eq:G1}
	\frac{dG_{x_1,x_2}}{dt}:={\mathcal L}^{(2)}(G_{x_1,x_2})= i(G_{x_1-1,x_2}+G_{x_1+1,x_2}-G_{x_1,x_2-1}-G_{x_1,x_2+1}) 
	\\-\gamma G_{x_1,x_2}(1-\delta_{x_1,x_2}) 
	-\frac{\gamma'}{2} G_{x_1,x_2}(\delta_{x_1,1}+\delta_{x_2,1}+\delta_{x_1,L}+\delta_{x_2,L})+\gamma'\delta_{x_1,1}\delta_{x_2,1}. 
\end{multline}
Eq.~\eqref{eq:G1} is the same as~\eqref{eq:G} apart for the boundary terms.  The boundary 
dissipation is modeled by the Lindblad operators $L_{x,2}=\sqrt{\gamma'} c_1^\dagger$  
and $L_{x,2}=\sqrt{\gamma'} c_L$. Importantly, Eq.~\eqref{eq:G1} becomes the same as~\eqref{eq:G} 
after the redefinition $\gamma^-=\gamma'/2$, apart for the ``driving'' term $\gamma'\delta_{x_1,1}\delta_{x_2,1}$. 
However, since Eq.~\eqref{eq:G} and~\eqref{eq:G1} are linear in $G_{x_1,x_2}$, given the general solution 
of~\eqref{eq:G},  it is possible to construct the solution of~\eqref{eq:G1} with generic initial condition 
$G_{x_1,x_2}^{\scriptscriptstyle(\mathrm{in})}$. Indeed, let us consider  
$ G^{\scriptscriptstyle (\mathrm{I})}_{x_1,x_2}$ solution of~\eqref{eq:G1} without the last term, and 
with initial condition $G_{x_1,x_2}(0)=G^{\scriptscriptstyle (\mathrm{in})}_{x_1,x_2}$. 
Let us also consider the solution $G^{\scriptscriptstyle (\mathrm{II})}_{x_1,x_2}$ of~\eqref{eq:G1} without 
the driving term and with delta initial condition $G^{\scriptscriptstyle(\mathrm{II})}_{x_1,x_2}(0)=\delta_{x_1,1}\delta_{x_2,1}$.  Now, 
one can verify that the solution of~\eqref{eq:G1} is 
\begin{equation}
	\label{eq:sol-gen}
	G_{x_1,x_2}(t)=G_{x_1,x_2}^{\scriptscriptstyle(\mathrm{I})}+
	\gamma'\int_0^t d\tau G^{\scriptscriptstyle (\mathrm{II})}_{x_1,x_2}(t-\tau). 
\end{equation}

In the following sections we will determine the full spectrum, i.e., the eigenvalues and the eigenvectors  
of ${\mathcal L}^{\scriptscriptstyle(2)}$ (cf.~\eqref{eq:G}), for arbitrary $\gamma,\gamma^-$ by using the Bethe ansatz. This 
allows us to  derive compact formulas for the fermionic correlator $G_{x_1,x_2}$ at arbitrary 
long times and chain sizes. In principle, by using~\eqref{eq:sol-gen} this also allows to obtain the dynamics of 
$G_{x_1,x_2}$ for the setup of Ref.~\cite{turkeshi2021diffusion}.

\section{Bethe Ansatz treatment of the Liouvillian ${\mathcal L}^{(2)}$}
\label{sec:ba}

\begin{figure}[t]
\centering
\includegraphics[width=.75\linewidth]{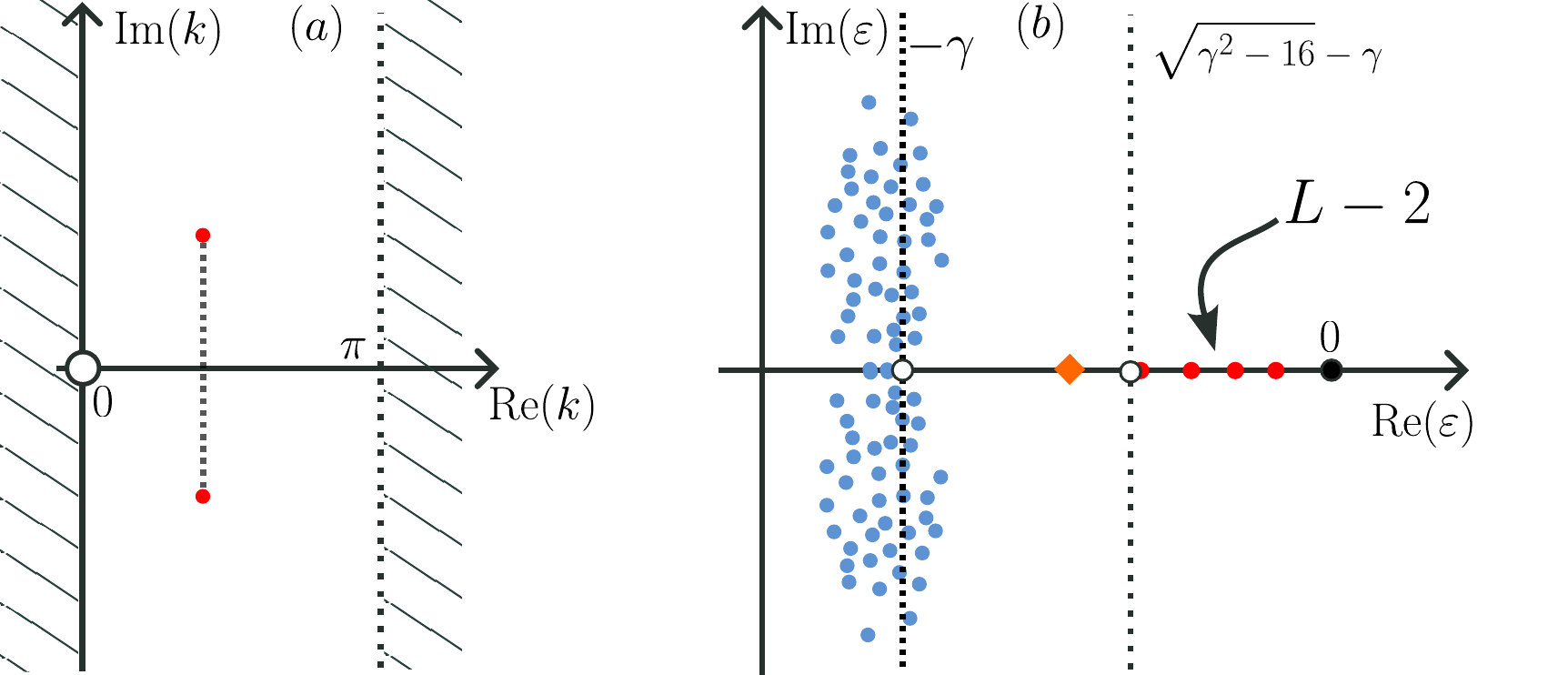}
\caption{Free-fermion chain with global dephasing and boundary losses. 
 (a) Allowed values for the solutions $k_1$ and $k_2$ of the 
 Bethe equations (cf.~\eqref{eq:beq-1} and~\eqref{eq:beq-2}). 
 Only the region $0<\mathrm{Re}(k)<\pi$ is allowed. Some of the 
 solutions form complex conjugate pairs as $k_1=k_2^*$, corresponding  
 to real eigenvalues $\varepsilon$. The remaining solutions 
 are all complex but do not form perfect strings. 
 (b) Typical structure of the Liouvillian spectrum: 
 $\mathrm{Im}(\varepsilon)$ versus $\mathrm{Re}(\varepsilon)$. 
 We only consider the situation with $\gamma>4$. The red circles are 
 purely real eigenvalues. They form an isolated diffusive band of eigenvalues.
 The diffusive band extends up to 
 $\varepsilon=\sqrt{\gamma^2-16}-\gamma$, and it contains $L$ eigenvalues 
 for generic $\gamma>4$ and $\gamma^-$. Upon lowering $\gamma$ the diffusive band 
 is depleted. The cluster around 
 $\varepsilon=-\gamma$ contains $\sim L(L-1)$ eigenvalues. 
 $L(L-1)/2$ of the levels are trivially obtained from the spectrum 
 of the model at $\gamma=0$. 
 The diamond denotes a pair of almost degenerate eigenvalues, which correspond 
 to boundary-localized modes of the Liouvillian. 
}
\label{fig1:cartoon}
\end{figure}

Here we discuss the Bethe ansatz framework that allows to solve~\eqref{eq:G}. We first 
introduce the ansatz \emph{\`a la} Bethe for the right eigenvectors of ${\mathcal L}^{\scriptscriptstyle(2)}$ 
in section~\ref{sec:bethe-states}. In section~\ref{sec:bethe-equation} we discuss the solutions of the 
Bethe equations and the general structure of the 
eigenvalues (energies) of the Liouvillian ${\mathcal L}^{\scriptscriptstyle(2)}$. In section~\ref{sec:deph-ind} 
we focus on a special class of states, which do not depend on dephasing, 
i.e., they are the same as in the tight-binding chain with 
boundary losses. In section~\ref{sec:vanishing} 
we discuss eigenvalues of ${\mathcal L}^{\scriptscriptstyle(2)}$ that are 
perturbatively related to the ones of section~\ref{sec:deph-ind} in the large $L$ limit. 
Finally, in section~\ref{sec:diff} we discuss solutions of the Bethe equations 
that form perfect strings in the complex plane (see Fig.~\ref{fig1:cartoon}). 
These states correspond to real eigenvalues and govern the long-time dynamics of the fermion correlator.

\subsection{Bethe ansatz for the eigenstates of ${\mathcal L}^{(2)}$} 
\label{sec:bethe-states}

Inspired by the coordinate Bethe Ansatz solution of the 
$XXZ$ chain with open boundary conditions~\cite{alcaraz1987surface} and 
by the Bethe Ansatz treatment of dephasing~\cite{esposito2005exactly,esposito2005emergence,medvedyeva2016exact} 
and incoherent hopping~\cite{eisler2011crossover} in free-fermion systems, 
we employ the following ansatz for $G_{x_1,x_2}$ as 
\begin{multline}
	\label{eq:ansatz}
	G_{x_1,x_2}=
\sum_{r_1,r_2=\pm} r_1 r_2 e^{\varepsilon(k_1,k_2) t}\Big\{\Big[A_{12}(r_1,r_2) e^{i r_1 k_1 x_1+i r_2 k_2 x_2}
	\\+(-1)^{x_1+x_2} A_{21}(r_1,r_2)e^{i r_2 k_1 x_2+i r_1 k_2 x_1}\Big]\Theta(x_2-x_1)
	+\sigma(-1)^{x_1+x_2}\Big[A_{12}(r_1,r_2) e^{i r_1 k_1 x_2+i r_2 k_2 x_1}\\
+(-1)^{x_1+x_2}A_{21}(r_1,r_2)e^{i r_2 k_1 x_1+i r_1 k_2 x_2}\Big]\Theta(x_1-x_2)\Big\}. 
\end{multline}
Here $k_1,k_2$ are complex quasimomenta, which have to be determined by solving the so-called 
Bethe equations. $G_{x_1,x_2}$ (after vectorization) are the right eigenvectors of 
${\mathcal L}^{\scriptscriptstyle (2)}$ with eigenvalues $\varepsilon(k_1,k_2)$. 
The evolution of~\eqref{eq:ansatz} is ``simple'' because ${\mathcal L^{\scriptscriptstyle(2)}} (G_{x_1,x_2})=\varepsilon G_{x_1,x_2}$, 
although it is not trivial, due to the eigenvalues $\varepsilon$ being complex. 
The sum over $r_1,r_2$ in~\eqref{eq:ansatz} is over the reflections of $k_1,k_2$, similar to the 
Bethe Ansatz solution of the Heisenberg chain~\cite{alcaraz1987surface} with boundary fields. 
The functions $\Theta(x)$ are Heaviside step functions. To recover the result for 
$x_1=x_2$, one has to take the limit $x_2=x_1+\epsilon$, sending 
$\epsilon>0$ to zero. This means that $G_{x_1,x_1}$ is given by the 
prefactor of the first Heaviside function in~\eqref{eq:ansatz}. The coefficients $A_{12}$ and $A_{21}$ are 
scattering amplitudes, which depend on $k_1,k_2$. 
Crucially, the Liouvillian ${\mathcal L}^{\scriptscriptstyle (2)}$ 
is invariant under the transformation ${\mathcal R}$  that transforms 
$G_{x_1,x_2}\to (-1)^{x_1+x_2} G_{x_2,x_1}$ as it 
can be verified by substitution in~\eqref{eq:G}. 
Since ${\mathcal R}^2$ is the identity, one has for the eigenfunctions of ${\mathcal R}$ that 
$(-1)^{x_1+x_2}G_{x_2,x_1}=\sigma G_{x_1,x_2}$, where $\sigma=\pm1$.  
The second term in~\eqref{eq:ansatz} takes into account this symmetry. 
%
%

To proceed, let us observe that in the bulk of the chain, 
i.e., for $1<x_1,x_2<L$, after substituting~\eqref{eq:ansatz} in~\eqref{eq:G}, we 
obtain the condition 
\begin{equation}
	\label{eq:bulk-eq}
	i(G_{x_1-1,x_2}+G_{x_1+1,x_2}-G_{x_1,x_2-1}-G_{x_1,x_2+1})-\gamma(1-\delta_{x_1,x_2})G_{x_1,x_2}-\varepsilon G_{x_1,x_2}=0. 
\end{equation}
Let us consider the situation with $x_1\ne x_2$. One can verify that the ansatz~\eqref{eq:ansatz} 
satisfies~\eqref{eq:bulk-eq} if we fix  
\begin{equation}
	\label{eq:eps}
	\varepsilon(k_1,k_2)=2i\cos(k_1)-2i\cos(k_2)-\gamma. 
\end{equation}
Importantly, the minus sign in the second term in~\eqref{eq:eps} depends on the 
choice of the ansatz~\eqref{eq:ansatz}. By redefining $k_2\to k_2+\pi$ 
in the terms that contain the sign factor $(-1)^{x_1+x_2}$ in~\eqref{eq:ansatz}, 
one obtains that $\varepsilon=2i\cos(k_1)+2i\cos(k_2)-\gamma$, 
which is symmetric under exchange $k_1\leftrightarrow k_2$ (see, for instance, Ref.~\cite{medvedyeva2016exact}). 
Notice that one has to change $k_2\to k_2+\pi$ also in the Bethe equations (see section~\ref{sec:bethe-equation}). 
After these redefinitions,  the new Bethe equations become the same as the Bethe equations 
for the Hubbard chain with imaginary boundary magnetic fields (see Appendix~\ref{app:hubb}). This happens  
despite the fact that the full Liouvillian contains a boundary pair production term (see Appendix~\ref{app:mapping}). 

Let us now determine the coefficients $A_{12}$ and $A_{21}$ in~\eqref{eq:ansatz}. It is convenient to 
to treat the cases $\sigma=1$ and $\sigma=-1$ separately. 
Let us start with the case with $\sigma=1$ in~\eqref{eq:ansatz}. As it will be clear in 
section~\ref{sec:deph-ind}, for $\sigma=-1$ the ansatz~\eqref{eq:ansatz} does not depend on 
$\gamma$, and the eigenstates of ${\mathcal L}^{\scriptscriptstyle (2)}$ are the same as those 
of the chain with boundary losses and no bulk dephasing. 
Let us now impose the ``contact'' condition obtained by fixing $x_1=x_2$ in~\eqref{eq:ansatz} and 
requiring that Eq.~\eqref{eq:bulk-eq} holds with $\varepsilon$ as in~\eqref{eq:eps}. 
A long calculation gives 
\begin{equation}
	\label{eq:cond-1}
	A_{21}(r_1,r_2)=-A_{12}(r_2,r_1)\frac{\gamma/2+r_2\sin(k_1)+r_1\sin(k_2)}{\gamma/2-r_2\sin(k_1)-r_1\sin(k_2)}. 
\end{equation}
Finally, we impose the boundary conditions. For Eq.~\eqref{eq:bulk-eq} to be 
compatible with  the Lindblad equation~\eqref{eq:G}   at the boundaries, we require that 
\begin{align}
	\label{eq:bound-1}
	& i G_{0,x_2}+\gamma^- G_{1,x_2}=0\\
	\label{eq:bound-2}
	& i G_{x_1,L+1}-\gamma^- G_{x_1,L}=0. 
\end{align}
The conditions~\eqref{eq:bound-1} and~\eqref{eq:bound-2} give eight equations. They 
allow us to fix 
\begin{align}
	& A_{12}(-1,1)=\frac{1-ie^{i k_1}\gamma^-}{1-ie^{-ik_1}\gamma^-}A_{12}(1,1)\\
	& A_{12}(1,-1)=e^{2ik_2(1+L)}\frac{1+ie^{-ik_2}\gamma^-}{1+ie^{ik_2}\gamma^-}A_{12}(1,1)\\
	& A_{12}(-1,-1)=e^{2ik_2(1+L)}\frac{1-ie^{i k_1}\gamma^-}{1-ie^{-ik_1}\gamma^-}
	\frac{1+ie^{-ik_2}\gamma^-}{1+ie^{ik_2}\gamma^-}A_{12}(1,1). 
\end{align}
Moreover, one obtains two more equations (Bethe equations) that provide the 
quantization conditions for $k_1$ and $k_2$. Before discussing the Bethe equations, 
let us stress that it is natural to expect that Eq.~\eqref{eq:bound-1} and~\eqref{eq:bound-2} 
can be modified to account for different loss rates $\gamma^-_L$ and $\gamma^-_R$ at the 
two edges of the chain. 

\subsection{Bethe equations and general structure of the Liouvillian spectrum}
\label{sec:bethe-equation}

The two extra conditions obtained from~\eqref{eq:bound-1} and~\eqref{eq:bound-2} 
provide two coupled nonlinear equations for $k_1,k_2$ as 
\begin{align}
	\label{eq:beq-1}
	& e^{2i k_1(L-1)}\Big(\frac{e^{ik_1}-i\gamma^-}{e^{-ik_1}-i\gamma^-}\Big)^2=
	\prod_{r_2=\pm1}\frac{\gamma/2-\sin(k_1)+r_2\sin(k_2)}{\gamma/2+\sin(k_1)-r_2\sin(k_2)}\\
	\label{eq:beq-2}
	& e^{2i k_2(L-1)}\Big(\frac{e^{ik_2}+i\gamma^-}{e^{-ik_2}+i\gamma^-}\Big)^2=
	\prod_{r_1=\pm1}\frac{\gamma/2-r_1\sin(k_1)+\sin(k_2)}{\gamma/2+r_1\sin(k_1)-\sin(k_2)}. 
\end{align}
Eq.~\eqref{eq:beq-1} and~\eqref{eq:beq-2} differ from the Bethe equations for the periodic 
tight-binding chain with dephasing. For periodic boundary conditions one has that $\gamma^-=0$ and 
one has to replace $e^{2i k_j L}\to e^{ik_j L}$. Moreover, only one of the two terms in the 
right-hand side survives, because there is no product over the reflections of the quasimomenta. 

Let us now discuss some properties of the Bethe equations~\eqref{eq:beq-1} and~\eqref{eq:beq-2}. 
The total number of solutions is $L^2$ because the Liouvillian ${\mathcal L}^{\scriptscriptstyle(2)}$ is 
a $L^2\times L^2$ matrix. The momenta $k_1,k_2$ are all complex. The allowed domain of $k_j$ is 
reported in Fig.~\ref{fig1:cartoon} (a), plotting $\mathrm{Im}(k_j)$ versus $\mathrm{Re}(k_j)$. 
The Bethe equations possess several symmetries that we now discuss. 
Given a generic pair of quasimomenta $(k_1,k_2)$ solving~\eqref{eq:beq-1} and~\eqref{eq:beq-2}, 
the pairs obtained by arbitrary reflections $\pm k_1$ and 
$\pm k_2$ are also solutions of~\eqref{eq:beq-1} and~\eqref{eq:beq-2}. This can be used to 
fix $\mathrm{Re}(k_j)>0$. Notice that $k_j=0$  and $k_j=\pi$ are solutions of the Bethe equations, although 
they have to be discarded because they give vanishing eigenvectors~\eqref{eq:ansatz}. 
The invariance of~\eqref{eq:beq-1} and~\eqref{eq:beq-2} under $k_1\to\pm k_2\pm\pi$  
can be exploited to fix $\mathrm{Re}(k_j)<\pi$. Since the imaginary part of $k_j$ can be 
arbitrary, the solutions of the Bethe equations live in the strip $(0,\pi)\times (-i\infty,i\infty)$ 
(see Fig.~\ref{fig1:cartoon} (a)). 
Another important property of~\eqref{eq:beq-1} and~\eqref{eq:beq-2} is that given a 
pair $(k_1,k_2)$ solving~\eqref{eq:beq-1}~\eqref{eq:beq-2}, then $(k_2^*,k_1^*)$, with the star denoting complex 
conjugation, is also a solution. This means that  (cf.~\eqref{eq:eps})  the eigenvalues $\varepsilon$ 
appear in complex conjugated pairs. 

Crucially, some of the solutions $(k_1,k_2)$ form complex conjugate pairs (see Fig.~\ref{fig1:cartoon} (a)), 
i.e., $k_1=k_2^*$. These solutions form ``strings'' patterns in the complex plane (see Fig.~\ref{fig1:cartoon}). 
The  corresponding eigenvalues are real. We anticipate that 
these solutions will determine the behavior of the fermionic correlator at long times, because  
the solutions giving the eigenvalues with the larger real parts will be of this type. 
String solutions of the Bethe equations can be effectively described in the limit $L\to\infty$ by using the framework of the string 
hypothesis~\cite{takahashi1999thermodynamics}. As we will discuss in section~\ref{sec:diff}, 
in the large $L$  limit the imaginary part of $k_1,k_2$ can be derived by solving a nonlinear equation, 
similar to the so-called Bethe-Gaudin-Takahashi (BGT) equation that appears within the 
framework of the string hypothesis for integrable models~\cite{takahashi1999thermodynamics}. 
Corrections to the string hypothesis  are exponentially suppressed in the limit $L\to\infty$. 

Let us now discuss the general structure  of the spectrum of ${\mathcal L}^{\scriptscriptstyle(2)}$. This is illustrated in 
Fig.~\ref{fig1:cartoon} (b). First, for $\gamma^-=0$, there is a zero-energy state $\varepsilon=0$, which 
corresponds to the steady-state of the system at $t\to\infty$. Within the Bethe ansatz treatment, 
the steady state corresponds to diverging momenta $k_1,k_2$. At nonzero $\gamma^-$, the eigenvalue 
$\varepsilon=0$ disappears. We can generically distinguish two different regions in the energy spectrum. 
The eigenvalues $\varepsilon$ with the larger real parts form a band of \emph{real}  energy near $\varepsilon=0$. 
As these solutions are responsible for diffusive behavior~\cite{esposito2005emergence,eisler2011crossover}, 
we dub them diffusive band~\cite{esposito2005emergence}. These are reminiscent of the diffusive band appearing  
for the periodic chain with dephasing~\cite{esposito2005emergence}, or incoherent hopping~\cite{eisler2011crossover}. 
The number of eigenvalues in the band depends on $\gamma$. As it will be clear in the following, 
for $\gamma>4$, the diffusive band contains $L$ eigenvalues at small $\gamma^-$, 
which is the largest possible number of states. 
Interestingly, for $\gamma^->\gamma^-_c=\exp(-\mathrm{arccosh}(\gamma/4))$ 
(see section~\ref{sec:diff}) two of the solutions move outside of the diffusive band towards lower $\mathrm{Re}(\varepsilon)$ 
(see the diamond symbol in Fig.~\ref{fig1:cartoon}). Concomitantly, the two eigenvalues  
become almost degenerate. Precisely, their splitting in 
energy decays exponentially with $L$. These states are boundary-related, and are present also in the 
one-dimensional Hubbard model with boundary fields~\cite{essler2005the}. 
Boundary-related states have been investigated 
in the two-particle sector for the open $XXZ$ spin chain in Ref.~\cite{alba2013Bethe}.

Furthermore, a cluster of eigenvalues is present around $\varepsilon=-\gamma$. As it will be clear in section~\ref{sec:deph-ind}, 
there are $L(L-1)/2$ eigenvalues that are related by a $\gamma$ shift to the eigenvalues of ${\mathcal L}^{\scriptscriptstyle(2)}$  
with $\gamma=0$, i.e., with only losses. Specifically, they 
are given by~\eqref{eq:ansatz} with $\sigma=-1$. The associated Bethe equations are decoupled and are given by~\eqref{eq:beq-3} 
and~\eqref{eq:beq-4}. The remaining complex eigenvalues correspond  states with $\sigma=1$ in~\eqref{eq:ansatz}. 
Still, in the large $L$  limit they differ from the states with $\sigma=-1$ by ${\mathcal O}(1/L)$ terms, i.e., they are 
``perturbatively'' related to the states with $\sigma=-1$. 

\subsection{Dephasing-independent solutions}
\label{sec:deph-ind}

Let us now characterize the states~\eqref{eq:ansatz} with $\sigma=-1$. Now, the main difference with the 
case $\sigma=1$ is that the contact condition~\eqref{eq:cond-1} has to be modified  as 
\begin{equation}
	A_{21}(r_1,r_2)=-A_{12}(r_2,r_1). 
\end{equation}
The boundary conditions to be imposed are the same as~\eqref{eq:bound-1} and~\eqref{eq:bound-2}.  
They give 
\begin{align}
	\label{eq:cd-1}
	& A_{12}(-1,1)=\frac{1-ie^{i k_1}\gamma^-}{1-ie^{-ik_1}\gamma^-}A_{12}(1,1)\\
	\label{eq:cd-2}
	& A_{12}(1,-1)=\frac{1+ie^{ik_2}\gamma^-}{1+ie^{-ik_2}\gamma^-}A_{12}(1,1)\\
	\label{eq:cd-3}
	& A_{12}(-1,-1)=\frac{1-ie^{i k_1}\gamma^-}{1-ie^{-ik_1}\gamma^-}
	\frac{1+ie^{ik_2}\gamma^-}{1+ie^{-ik_2}\gamma^-}A_{12}(1,1). 
\end{align}
Notice that there is no dependence on $L$ in~\eqref{eq:cd-1}-\eqref{eq:cd-3}, 
in contrast with the case with $\sigma=1$. The Bethe equations now read as 
\begin{align}
	\label{eq:beq-3}
	&(\gamma^-)^2\sin(k_2(1-L))+2i\gamma^-\sin(k_2 L)+\sin(k_2(L+1))=0\\ 
	\label{eq:beq-4}
	&(\gamma^-)^2\sin(k_1(1-L))-2i\gamma^-\sin(k_1 L)+\sin(k_1(L+1))=0. 
\end{align}
The Bethe equations for $k_1$ and $k_2$ are decoupled, reflecting that the 
system is noninteracting. Also, $k_1,k_2$ do not 
depend on the dephasing rate $\gamma$. The eigenvalues $\varepsilon$ are the same as 
in~\eqref{eq:eps}, implying that the dependence on $\gamma$ is only a shift. As it is clear 
from~\eqref{eq:beq-4}, given a solution $k_1$, then $-k_1$ and $k_1\pm\pi$ is also a 
solution. The same holds for $k_2$. This means that one can restrict to $0<\mathrm{Re}(k_j)<\pi$. 
Moreover, the solutions of Eq.~\eqref{eq:beq-3} and Eq.~\eqref{eq:beq-4} are related by 
complex conjugation. 

Eq.~\eqref{eq:beq-3} and~\eqref{eq:beq-4} are 
the same equations describing the tight-binding chain with boundary losses and no 
bulk dephasing~\cite{guo2018analytical}. Since Eq.~\eqref{eq:beq-3} has $L$ solutions $k_1^{\scriptscriptstyle(p)}$ 
with $p=1,\dots, L$, 
the pairs $(k^{\scriptscriptstyle (p)}_1,(k^{\scriptscriptstyle(q)}_1)^*)$ 
give all the $L^2$ eigenvalues of the Liouvillian. Upon switching on $\gamma$, only $L(L-1)/2$ survive. 
These correspond to the pairs $(k_1,k_2)$ such that $k_j\ne0,\pi$. Moreover, one has to exclude all the 
pairs $k_1,k_2$ such that $k_1+k_2=0\mod \pi$, and the pairs $(k_1',k_2')$ that 
are obtained as  $(k_1',k_2')=(\pi-k_2,\pi-k_1)$ from a set of solutions $(k_1,k_2)$. 
Indeed, one can check that the total number of pairs satisfying these constraints is $L(L-1)/2$. 
The conditions on $k_1,k_2$ discussed above are the same as in the beginning of Section~\ref{sec:bethe-equation}. 
Let us also observe that in the limit $L\to\infty$ the solutions of~\eqref{eq:beq-3} and~\eqref{eq:beq-4} 
are given as 
\begin{equation}
	\label{eq:kr}
	k_{1,2}=\frac{\pi}{L+1}j_{1,2}+{\mathcal O}(1/L), \quad j_{1,2}=1,2,\dots L.  
\end{equation}
Specifically, the imaginary part of $k_{1,2}$ is ${\mathcal O}(1/L)$, although it is nonzero. 
Clearly, Eq.~\eqref{eq:kr} is exact without the ${\mathcal O}(1/L)$ correction 
for $\gamma^-=0$. In the last case the eigenvalues $\varepsilon$ form  
a straight line parallel to the imaginary axis (see Fig.~\ref{fig1:cartoon}), 
with real part $-\gamma$.

\subsection{Solutions  with vanishing imaginary parts of $k_1$ and $k_2$}
\label{sec:vanishing}

Near the eigenvalues that correspond to dephasing-independent solutions of the Bethe equations, 
there are $\sim L(L-1)/2$ eigenvalues that correspond to $\sigma=1$ in~\eqref{eq:ansatz}, and that 
in the large $L$ limit  differ by terms ${\mathcal O}(1/L)$ from the dephasing-independent eigenvalues. 
The number of eigenvalues depends on $\gamma$. In particular, for $\gamma>4$ their number is exactly $L(L-1)/2$. 

We now discuss them restricting to the case with $\gamma^-=0$. A similar analysis can be performed for nonzero $\gamma^-$. 
The large $L$ behavior of the Bethe equations~\eqref{eq:beq-1}~\eqref{eq:beq-2} suggests the expansion 
\begin{align}
	\label{eq:k1-L}
	&k_1=k_1^{(r,0)}+k_{1}^{(r,2)}L^{-2}+i k_1^{(i,1)}L^{-1}\\
	\label{eq:k2-L}
	&k_2=k_2^{(r,0)}+k_{2}^{(r,2)}L^{-2}+i k_2^{(i,1)}L^{-1}. 
\end{align}
Here $k_j^{\scriptscriptstyle(r,0)},k_j^{\scriptscriptstyle(r,2)}$ and $k_j^{\scriptscriptstyle(i,1)}$ 
($j=1,2$) are real parameters that have to be determined. 
After  substituting the 
ansatz~\eqref{eq:k1-L} and~\eqref{eq:k2-L} in the Bethe equations~\eqref{eq:beq-1} and~\eqref{eq:beq-2}, 
Taylor expanding in the large $L$ limit, and equating the coefficients of the terms with the same powers of $L$, 
we obtain 
\begin{equation}
	\label{eq:string-trivial}
	2k^{(r,0)}_1(L+1)= j_1\pi,\quad 2k_2^{(r,0)}(L+1)=j_2\pi, \quad\mathrm{with}\,\,j_1,j_2=1,2,\dots 2(L+1). 
\end{equation}
We now provide the expression for $k_j^{\scriptscriptstyle(i,1)}$. A similar expression can be obtained 
for $k_j^{\scriptscriptstyle(r,2)}$, although since it is cumbersome we do not report it. 
We obtain 
\begin{align}
	\label{eq:k1-large-L}
	& k_1^{(i,1)}=-\frac{1}{2}\frac{L}{L+1}\ln\left[(-1)^{j_1}
	\frac{\gamma^2-2\cos(2k_1^{(r,0)})+2\cos(2k_2^{(r,0)})-4\gamma\sin(k_1^{(r,0)})}{
\gamma^2-2\cos(2k_1^{(r,0)})+2\cos(2k_2^{(r,0)})+4\gamma \sin(k_1^{(r,0)})}\right]\\
\label{eq:k2-large-L}
	& k_2^{(i,1)}=-\frac{1}{2}\frac{L}{L+1}\ln\left[(-1)^{j_2}
	\frac{\gamma^2+2\cos(2k_1^{(r,0)})-2\cos(2k_2^{(r,0)})+4\gamma\sin(k_2^{(r,0)})}{
\gamma^2+2\cos(2k_1^{(r,0)})-2\cos(2k_2^{(r,0)})-4\gamma \sin(k_2^{(r,0)})}\right]. 
\end{align}
Consistency with~\eqref{eq:k1-L} and~\eqref{eq:k2-L} requires that  $k_j^{\scriptscriptstyle(i,1)}$ is 
real. One can readily check that for $\gamma>4$ the term inside the square brackets in~\eqref{eq:k1-large-L} and~\eqref{eq:k2-large-L} 
is positive provided that $j_1$ and $j_2$ are both even. For $\gamma<4$, Eq.~\eqref{eq:k1-large-L} and~\eqref{eq:k2-large-L} 
are correct only for the eigenvalues near $\varepsilon= -\gamma$. Oppositely, away from $\varepsilon=-\gamma$ 
the eigenvalues are affected by the presence of the diffusive band, and are not accurately 
described by~\eqref{eq:k1-large-L} and~\eqref{eq:k2-large-L}. 
Notice that $k_{1,2}^{\scriptscriptstyle(i,1)}$ vanish in the large $L$ limit, even at $j_{1,2}/L$ 
fixed. 
%
\begin{figure}[t]
\centering
\includegraphics[width=.75\linewidth]{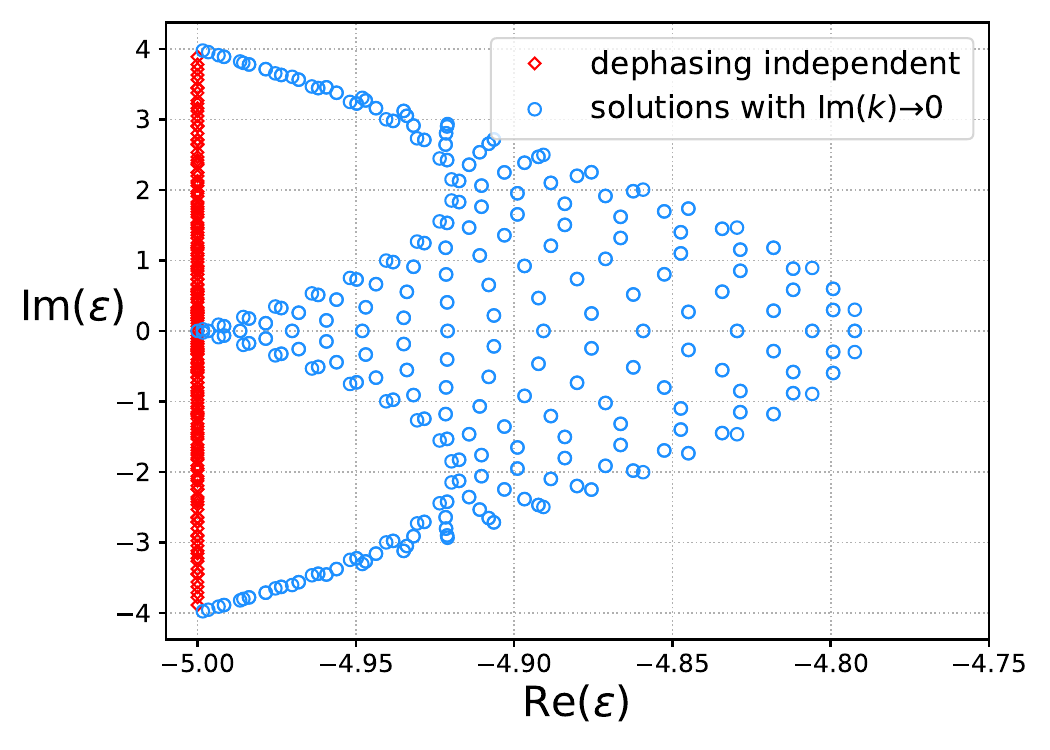}
\caption{ Spectrum of ${\mathcal L}^{(2)}$ for $L=20$, $\gamma=5$ 
 and $\gamma^-=0$. We plot $\mathrm{Im}(\varepsilon)$ versus 
 $\mathrm{Re}(\varepsilon)$. We only show the eigenvalues near $\varepsilon=-\gamma$. 
 The circles correspond to the 
 complex solutions $(k_1,k_2)$ of the Bethe equations with vanishing 
 imaginary parts $\mathrm{Im}(k_j)\to0$ in the limit $L\to\infty$ (see section~\ref{sec:vanishing}). 
 The diamonds are the solutions that do not depend on the dephasing rate $\gamma$ (see section~\ref{sec:deph-ind}). 
}
\label{fig:vanishing}
\end{figure} 
%
The eigenvalues $\varepsilon$ that correspond to solutions of the Bethe equations with vanishing 
imaginary parts are discussed in Fig.~\ref{fig:vanishing}. We consider only the case with 
$\gamma=5$ because for $\gamma>4$ the diffusive band at smaller $\mathrm{Re}(\varepsilon)$ is well 
separated from bulk of the spectrum and~\eqref{eq:k1-L} and~\eqref{eq:k2-L} are accurate. The diamonds in the Figure 
correspond to the eigenstates with $\sigma=-1$ in~\eqref{eq:ansatz} discussed in section~\ref{sec:deph-ind}. 
The circles are the eigenvalues obtained from to momenta of the type~\eqref{eq:k1-L} and~\eqref{eq:k2-L}. 
It is interesting to focus on the inner and outer ``envelope'' of the eigenvalues. 
They are obtained 
from~\eqref{eq:string-trivial}~\eqref{eq:k1-large-L} and~\eqref{eq:k2-large-L} as 
follows. We checked that the levels $\varepsilon_\mathrm{in}$ of the inner envelope 
correspond to the choice $j_1=2$ and $j_2\in[2,2(L+1)]$ and $j_1\in[2,2(L+1)]$ in~\eqref{eq:string-trivial}. 
The levels $\varepsilon_\mathrm{out}$ forming the outer envelope are 
obtained by choosing $j_1\in [2,2(L+1)]$ and $j_2=2(L+1)-j_1$.

\subsection{Diffusive band \& boundary states}
\label{sec:diff}

As we anticipated, the eigeneigenvalues of ${\mathcal L}^{\scriptscriptstyle(2)}$ having the largest real part form a 
diffusive band, and are real. These states correspond to string solutions of the Bethe equations (see Fig.~\ref{fig1:cartoon} (a)) 
with nonvanishing imaginary parts in the limit $L\to\infty$. 
They form complex conjugate pairs $(k_1,k_1^*)$. It is convenient to define 
$k_1=k_{r}+i k_{i}$ and $k_2=k_{r}-i k_i$. We start discussing the case with $\gamma^-=0$. 
In terms of $k_r,k_i$, the Bethe equations~\eqref{eq:beq-1} and~\eqref{eq:beq-2}  become 
\begin{align}
	\label{eq:baer}
	& e^{2(L+1)(ik_r- k_i)}=\frac{(\gamma-4\cosh(k_i)\sin(k_r))(\gamma-4i\cos(k_r)\sinh(k_i))}
	{(\gamma+4\cosh(k_i)\sin(k_r))(\gamma+4i\cos(k_r)\sinh(k_i))}\\
	\label{eq:baei}
	& e^{2(L+1)(ik_r+ k_i)}=\frac{(\gamma+4\cosh(k_i)\sin(k_r))(\gamma-4i\cos(k_r)\sinh(k_i))}
	{(\gamma-4\cosh(k_i)\sin(k_r))(\gamma+4i\cos(k_r)\sinh(k_i))}. 
\end{align}
To proceed,  we can assume without loss of generality that $k_i>0$. In the limit $L\to\infty$ the left-hand side 
of~\eqref{eq:baei} diverges exponentially. This suggests that the denominator in the right-hand side 
of~\eqref{eq:baer} vanishes. For consistency we can impose that 
\begin{equation}
	\label{eq:impo}
	\gamma-4\cosh(k_i)\sin(k_r)=0. 
\end{equation}
Similar observations are at the heart of the string hypothesis in Bethe ansatz solvable 
models~\cite{takahashi1999thermodynamics}. 
Solving~\eqref{eq:impo} for $k_r$, we obtain 
\begin{equation}
	\label{eq:string}
	k_r=\pi-\arcsin\left(\frac{\gamma}{4\cosh(k_i)}\right). 
\end{equation}
To remove the singular denominator in~\eqref{eq:baer} we 
take the product of~\eqref{eq:baer} and~\eqref{eq:baei}, and 
after using~\eqref{eq:string}, we obtain that $k_i$ satisfies the equation 
\begin{equation}
	\label{eq:bgt-0}
	\left(\frac{4}{\gamma\sech(k_i)-i\sqrt{16-\gamma^2\sech^2(k_i)}}\right)^{4(L+1)}
		=\left(\frac{\gamma\csech(k_i)+i\sqrt{16-\gamma^2\sech^2(k_i)}
		}{\gamma\csech(k_i)-i\sqrt{16-\gamma^2\sech^2(k_i)}}\right)^2. 
\end{equation}
The derivation of~\eqref{eq:bgt-0} is similar to that of the Bethe-Gaudin-Takahashi (BGT) 
equations for  the Hubbard chain~\cite{essler2005the,takahashi1999thermodynamics}. 
For this reason, we refer to~\eqref{eq:bgt-0} as the BGT equation. 
The derivation  can be extended to the case with nonzero $\gamma^-$. 
The BGT equation becomes 
\begin{multline}
	\label{eq:bgt}
	\frac{(4e^{k_i}+\gamma\gamma^-\sech(k_i)+i\gamma^-{\mathcal B}(k_i))^2
	(4e^{-k_i}-\gamma\gamma^-\sech(k_i)-i\gamma^-{\mathcal B}(k_i))^2}{
		(4e^{k_i}\gamma^--\gamma\sech(k_i)-i{\mathcal B}(k_i))^2
	(4e^{-k_i}\gamma^-+\gamma\sech(k_i)+i{\mathcal B}(k_i))^2}	\\
	\times     \frac{(\gamma\csech(k_i)+i{\mathcal B}(k_i))^2}{
	(\gamma\csech(k_i)-i{\mathcal B}(k_i))^2} -
	\left(\frac{4}{\gamma\sech(k_i)-i{\mathcal B}(k_i)}\right)^{4L}=0, 
\end{multline}
where we defined 
\begin{equation}
	{\mathcal B}(x):=\sqrt{16-\gamma^2\sech^2(x)}. 
\end{equation}
After solving~\eqref{eq:bgt} for $k_i$, we obtain $k_r$ by using~\eqref{eq:string}. 
It is convenient to take the logarithm of both terms in~\eqref{eq:bgt} to obtain 
the BGT equations in logarithmic form. Let us first define 
\begin{equation}
	\label{eq:z-def}
	z:=\frac{\gamma}{4\cosh(k_{i})},\quad k_i=\mathrm{arccosh}\left(\frac{\gamma}{4z}\right). 
\end{equation}
After taking the logarithm of both members in~\eqref{eq:bgt} and using~\eqref{eq:z-def}, 
we obtain 
\begin{multline}
	\label{eq:bgt-eq}
	2iL\arcsin(z_j)+\ln\left[\frac{z_j\gamma+i\sqrt{1-z_j^2}\sqrt{\gamma^2-16z_j^2}}{z\gamma-i\sqrt{1-z_j^2}\sqrt{\gamma^2-16z_j^2}}\right]
	\\+\ln\left[\frac{\gamma^-(-i z_j+\sqrt{1-z_j^2})(\gamma+\sqrt{\gamma^2-16z_j^2})+4iz_j}{(iz_j-
	\sqrt{1-z_j^2})(\gamma+\sqrt{\gamma^2-16z_j^2})+4iz_j\gamma^-}\right]\\
	+\ln\left[\frac{\gamma+\sqrt{\gamma^2-16z_j^2}+4z_j(z_j+i\sqrt{1-z_j^2})
	\gamma^-}{4z_j(z_j+i\sqrt{1-z_j^2})-(\gamma+\sqrt{\gamma^2-16z_j^2})
\gamma^-}\right]=-\pi i I_j.
\end{multline}
Here $I_j\in[0,L-2)$ are integers, forming 
the so-called BGT quantum numbers, which identify the different solutions $z_j$.  
They originate from the branch cut of the logarithm. The energy with the largest real 
part corresponds to $I_j=0$. Notice that here we assume $\gamma^->0$. For $\gamma^-=0$ 
$I_j=0$ has to be excluded, because it would correspond to $\varepsilon=0$. 
The number of solutions in the diffusive band depends on 
$\gamma$ and $\gamma^-$. Specifically, for $\gamma>4$ there are  at least $L-2$ solutions 
of~\eqref{eq:bgt-eq}. Two extra solutions appear provided that $\gamma^-$ is small enough. 
Precisely, for $\gamma^->\gamma^-_c$ two of the eigenvalues detach from the diffusive band, 
moving towards lower $\mathrm{Re}(\varepsilon)$. They correspond to boundary-related modes of the Liouvillian ${\mathcal L}^{\scriptscriptstyle(2)}$ 
(see diamond symbol in Fig.~\ref{fig1:cartoon}). 
Similar boundary states appear in the open Hubbard chain with boundary magnetic fields~\cite{essler2005the}. 
Precisely, from~\eqref{eq:bgt} we obtain that the boundary-related states are present  
for $\gamma^->\gamma^-_c$ given as 
\begin{equation}
	\label{eq:condition}
	\gamma^-_c=\exp\left(-\mathrm{arccosh}(\gamma/4)\right). 
\end{equation}
The condition~\eqref{eq:condition} is obtained by noticing that at the left edge of the diffusive band 
(see Fig.~\ref{fig1:cartoon}) one has that $k_\mathrm{i}=\mathrm{arccosh}(\gamma/4)$, and by solving 
Eq.~\eqref{eq:bgt} for $\gamma^-$. Eq.~\eqref{eq:condition} holds true for $\gamma>4$, and, since 
we are employing the framework of the string hypothesis, in the thermodynamic limit $L\to\infty$. 
Finally, we should stress that to extract the quasimomenta that correspond to the boundary-related 
eigenvalues it is convenient to use Eq.~\eqref{eq:bgt} rather than the logarithmic BGT equation~\eqref{eq:bgt-eq}, 
as it will be  clear in section~\ref{sec:bgt-eq} (see Fig.~\ref{fig:bethe-num-1}). 

Let us now discuss the gap of the Liouvillian ${\mathcal L}^{\scriptscriptstyle(2)}$. This is obtained 
by considering the energy $\varepsilon$ with the largest nonzero 
real part. Precisely, the gap $\Delta{\mathcal L}^{\scriptscriptstyle(2)}$ is defined as 
\begin{equation}
	\label{eq:gap-def}
	\Delta{\mathcal L}^{\scriptscriptstyle(2)}:=-\max_{j}\mathrm{Re}(\varepsilon_j),\quad\mathrm{with}\,\,
	\mathrm{Re}(\varepsilon_j)\ne0. 
\end{equation}
We numerically verified that for nonzero $\gamma^-$, $\Delta{\mathcal L}^{\scriptscriptstyle(2)}$ corresponds 
to $I_j=0$ in~\eqref{eq:bgt-eq}. For $\gamma^-=0$ one has to choose $I_j=1$. 
Focusing on $\gamma^-\ne0$, a straightforward expansion of~\eqref{eq:bgt-eq} for $I_j=0$ in the large $L$ 
limit gives 
\begin{equation}
	z\simeq\frac{\pi}{2}\frac{1}{L}+\left(\frac{\pi}{2}-\frac{\pi(1-(\gamma^-)^2)}{\gamma\gamma^-}
	\right)\frac{1}{L^2}, 
\end{equation}
where we neglected higher order terms in $1/L$. 
After substituting in the expression for the energy~\eqref{eq:eps} we obtain the 
gap of the Liouvillian as 
\begin{equation}
	\label{eq:liou-gap}
	\Delta{\mathcal L}^{(2)}\simeq-\frac{2\pi^2}{\gamma L^2}+\frac{4\pi^2(2-\gamma\gamma^-
	+2(\gamma^-)^2)}{\gamma^2\gamma^-}\frac{1}{L^3}. 
\end{equation}
At the leading order in $1/L$, the gap depends only on the bulk dephasing. At 
higher orders a dependence on $\gamma^-$ appears. This reflects that the 
effect of the boundaries appear at later times.

\section{Asymptotic dynamics of the fermionic two-point function}
\label{sec:asy}

Here we derive a formula for the time-dependent correlation function $G_{x_1,x_2}(t)$ starting 
from an arbitrary initial condition $G_{x_1,x_2}(0)$. The strategy is to build a complete basis 
of operators by using the Bethe states~\eqref{eq:ansatz}. This basis is then used to expand the 
initial condition for the correlator. In section~\ref{sec:left} we 
construct the complete basis for the generic two-point correlation function, using the left and right 
eigenvectors of ${\mathcal L}^{\scriptscriptstyle(2)}$. In section~\ref{sec:norm} we compute the 
leading contribution of the norm of the Bethe states~\eqref{eq:ansatz} in the large $L$ limit. We only consider the 
states forming the diffusive band, because they are dominant in the long-time limit. 
Finally, in section~\ref{sec:green} we derive the long-time limit of the density profile, i.e., the 
diagonal correlator $G_{x,x}(t)$.

\subsection{Left and right eigenvectors of ${\mathcal L}^{(2)}$}
\label{sec:left}

One can decompose the initial correlator $G_{x_1,x_2}(0)$ in the basis 
of eigenvectors of ${\mathcal L}^{\scriptscriptstyle (2)}$.  
Let us denote with $G_{x_1,x_2}^{\scriptscriptstyle(k_1,k_2)}$ the eigenvector of 
${\mathcal L}^{\scriptscriptstyle(2)}$ identified by the 
solutions $k_1,k_2$ of the Bethe equations~\eqref{eq:beq-1}~\eqref{eq:beq-2}. 
The dynamics of $G_{x_1,x_2}^{\scriptscriptstyle(k_1,k_2)}$ is given as 
\begin{equation}
	\frac{d G^{(k_1,k_2)}_{x_1,x_2}}{dt}={\mathcal L}^{(2)}(G^{(k_1,k_2)}_{x_1,x_2})=
	\varepsilon(k_1,k_2)G_{x_1,x_2}^{(k_1,k_2)}. 
\end{equation}
The generic correlator $G_{x_1,x_2}$ can be decomposed  as 
\begin{equation}
	\label{eq:G-exp}
	G_{x_1,x_2}(t)=\sum_{\{k_1,k_2\}} 
	|k_1,k_2\rangle\langle k_1,k_2|0\rangle 
	e^{\varepsilon(k_1,k_2) t}, \quad |k_1,k_2\rangle:=N_{k_1,k_2}^{-1}G_{x_1,x_2}^{(k_1,k_2)}. 
\end{equation}
where the sum is over the solutions of the Bethe equations $\{k_1,k_2\}$ 
(cf.~\eqref{eq:beq-1} and~\eqref{eq:beq-2}). In~\eqref{eq:G-exp} we redefined 
$|0\rangle:=G_{x_1,x_2}(0)$, and we defined 
$\langle k_1,k_2|:=\bar G_{x_1,x_2}^{\scriptscriptstyle (k_1,k_2)}$ as  the left eigenvector 
of the Liouvillian. Since the Liouvillian is not hermitian 
we have that $\langle k_1,k_2|\ne (|k_1,k_2\rangle)^\dagger$. 
In~\eqref{eq:G-exp} we defined the ``scalar product'' 
\begin{equation}
	\label{eq:scalar}
	\langle k_1,k_2|k_3,k_4\rangle:=\sum_{z_1,z_2=1}^L \bar 
	G_{z_1,z_2}^{(k_1,k_2)}G_{z_1,z_2}^{(k_3,k_4)}. 
\end{equation}
Similar definition holds for the scalar product with the initial correlator $|0\rangle$, 
i.e., $\langle k_1,k_2|0\rangle$. In~\eqref{eq:G-exp} 
$N_{k_1,k_2}:=\langle k_1,k_2|k_1,k_2\rangle$ is the normalization of the eigenvectors. 
Following Ref.~\cite{eisler2011crossover}, it is possible to determine 
$\langle k_1,k_2|$ by observing that the energy~\eqref{eq:eps} is invariant under the redefinition 
\begin{equation}
	\label{eq:kredef}
	k_1=\pi-k_2,\quad k_2=\pi-k_1. 
\end{equation}
This leads to the definition
\begin{equation}
\label{eq:dagger}
\langle k_1,k_2|:=|\pi-k_2,\pi-k_1\rangle. 
\end{equation}
Under the redefinition~\eqref{eq:kredef}, the ansatz~\eqref{eq:ansatz} gets modified. 
Apart from the redefinition~\eqref{eq:kredef}, 
one obtains new scattering amplitudes $\bar A_{12}$ and $\bar A_{21}$. For $\sigma=1$ 
they read as 
\begin{equation}
	\label{eq:S12left}
 \bar A_{21}(r_1,r_2)=-\bar A_{12}(r_2,r_1)\frac{\gamma/2+r_1\sin(k_1)+r_2\sin(k_2)}{\gamma/2-r_1\sin(k_1)-r_2\sin(k_2)}. 
\end{equation}
Notice that~\eqref{eq:S12left} differs from~\eqref{eq:cond-1} by the 
exchange $r_1\leftrightarrow r_2$ in the 
ratio on the right-hand side. Moreover, we have that 
\begin{align}
	& \bar A_{12}(-1,1)=\frac{1+ie^{-ik_2}\gamma^-}{1+ie^{ik_2}\gamma^-}\bar A_{12}(1,1)\\
	& \bar A_{12}(1,-1)=e^{-2ik_1(L+1)}\frac{1-ie^{ik_1}\gamma^-}{1-ie^{-ik_1}\gamma^-}A_{12}(1,1)\\
	& \bar A_{12}(-1,-1)=
	e^{-2ik_1(L+1)}\frac{1-ie^{ik_1}\gamma^-}{1-ie^{-ik_1}\gamma^-}
	\frac{1+ie^{-ik_2}\gamma^-}{1+ie^{ik_2}\gamma^-}\bar A_{12}(1,1). 
\end{align}
Importantly, the Bethe equations remain the same as in~\eqref{eq:beq-1} and~\eqref{eq:beq-2}. 

Let us now discuss the case of $\sigma=-1$. Repeating the steps as above we obtain that 
\begin{equation}
	\bar A_{21}(r_1,r_2)=-\bar A_{12}(r_2,r_1). 
\end{equation}
Moreover, we have 
\begin{align}
	& \bar A_{12}(-1,1)=\frac{1+ie^{-i k_2}\gamma^-}{1+ie^{ik_2}\gamma^-}\bar A_{12}(1,1)\\
	& \bar A_{12}(1,-1)=\frac{1-i e^{-i k_1}\gamma^-}{1-ie^{ik_1}\gamma^-}\bar A_{12}(1,1,)\\
	& \bar A_{12}(-1,-1)=\bar A_{12}(1,-1)\bar A_{12}(-1,1)/\bar A_{12}(1,1). 
\end{align}
We anticipate, however, that since the states with $\sigma=-1$ have typically small ${\mathrm{Re}}(\varepsilon)$, 
they do not contribute significantly at long times. Moreover, for delta initial 
conditions $G_{x_1,x_2}(0)$ considered in section~\ref{sec:green} their contribution is exactly zero. 

\subsection{Norm of the  eigenvectors}
\label{sec:norm}

Here we derive the norm $N_{k_1,k_2}$ for a Bethe eigenstate~\eqref{eq:ansatz} characterized 
by generic solutions of the Bethe equations $k_1,k_2$. 
To proceed, let us focus on eigenstates with $\sigma=1$ (cf.~\eqref{eq:ansatz}) because they 
dominate the dynamics at long times. By using~\eqref{eq:ansatz}~\eqref{eq:scalar} and~\eqref{eq:dagger} 
we obtain that 
\begin{equation}
	N_{k_1,k_2}=\langle k_1,k_2|k_1,k_2\rangle=N_0+N_1 L+N_2L^2, 
\end{equation}
where $N_0,N_1,N_2$ are functions of $k_1,k_2$. 
A tedious calculation gives 
\begin{equation}
	\label{eq:norm1}
	N_2=-8\frac{\gamma/2+\sin(k_1)+\sin(k_2)}{\gamma/2-\sin(k_1)-\sin(k_2)}. 
\end{equation}
A similar calculation gives $N_1$ as 
\begin{multline}
	\label{eq:norm2}
N_1=4i\gamma(\cos(k_1)-\cos(k_2))\Big[\frac{1}{(\gamma/2-\sin(k_1)-\sin(k_2))^2}\\
+\frac{\gamma/2+\sin(k_1)+\sin(k_2)}{(\gamma/2-\sin(k_1)-\sin(k_2))(\gamma/2-\sin(k_1)+\sin(k_2))(\gamma/2+\sin(k_1)-\sin(k_2))}\Big]\\
-\frac{\gamma/2+\sin(k_1)+\sin(k_2)}{\gamma/2-\sin(k_1)-\sin(k_2)}\Big[
\frac{8(1+(\gamma^-)^2)}{(1-i\gamma^-e^{ik_1})(1-i\gamma^- e^{-ik_1})}\\+
\frac{8(1+(\gamma^-)^2)}{(1+i\gamma^-e^{ik_2})(1+i\gamma^- e^{-ik_2})}\Big]. 
\end{multline}
Finally, we observe that $N_0$ is in general nonzero. This is in contrast with the case of the tight-binding 
chain with  incoherent hopping and periodic boundary conditions~\cite{eisler2011crossover}. 
Crucially, we notice that for the string solutions of the Bethe equations forming the 
diffusive band (see section~\ref{sec:diff}) both $N_1$ and $N_2$ are singular in the  limit 
$L\to\infty$ because $\gamma/2-\sin(k_1)-\sin(k_2)$ vanishes. Similar divergences will plague 
the overlaps $\langle k_1,k_2|0\rangle$ and $|k_1,k_2\rangle$. This means that to extract 
the dynamics of $G_{x_1,x_2}$ one has to take carefully the limit $L\to\infty$, i.e., going 
beyond the string hypothesis. In the following section we will show that to obtain 
the leading behavior of the norm in the large $L$ limit it is sufficient to consider the 
term in~\eqref{eq:norm2}. 

\subsection{Evolution of the density profile}
\label{sec:green}

In this section we provide analytic results for the dynamics of $G_{x,x}(t)$ starting from 
the initial condition 
\begin{equation}
	\label{eq:initial}
	|0\rangle:=G_{x_1,x_2}(0)=\delta_{x_1,x}\delta_{x_2,x}, 
\end{equation}
which corresponds to a fermion initially localized at position $x$. 
Our results hold in the long-time limit and for large $L$. 

Crucially, since Eq.~\eqref{eq:G}  
is linear as a function of $G_{x_1,x_2}$, its solution with the delta initial condition~\eqref{eq:initial} 
is sufficient to obtain the dynamics of $G_{x_1,x_2}$ starting from an arbitrary initial 
density profile. Precisely, let us consider a generic diagonal initial condition as 
\begin{equation}
	\label{eq:initial-1}
	|0\rangle=\delta_{x_1,x_2}f(x_2), 
\end{equation}
where $0\le f(x)\le1$. Given the solution $G^{\scriptscriptstyle (x)}_{x_1,x_2}(t)$ with  
initial condition~\eqref{eq:initial} at fixed $x$, the solution $G_{x_1,x_2}(t)$ with 
initial condition~\eqref{eq:initial-1} is obtained as 
\begin{equation}
	G_{x_1,x_2}(t)=\sum_{x=1}^L G_{x_1,x_2}^{\scriptscriptstyle(x)}(t)f(x). 
\end{equation}

Let us now discuss the time-dependent correlator $G_{x_1,x_2}(t)$ starting from~\eqref{eq:initial}. 
 To be specfic, let us consider the situation with a 
fermion initially localized at $x$ away from the boundaries, i.e., with $x/L\ne 0,1$ 
in~\eqref{eq:initial}. To obtain  $G_{x_1,x_2}(t)$ we  employ~\eqref{eq:G-exp}, 
restricting ourselves to the eigenstates of the Liouvillian  forming the diffusive 
band (see section~\ref{sec:diff}). 
Moreover, we exploit the string 
hypothesis, which holds in the limit $L\to\infty$. 
A straightforward calculation gives the overlaps between~\eqref{eq:initial} 
and the generic Bethe eigenstate~\eqref{eq:ansatz} as 
\begin{equation}
	\label{eq:overlap}
	\langle k_1,k_2|0\rangle=
	\sum_{r_1,r_2=\pm} r_1 r_2 \left[e^{-i (r_2 k_1+r_1 k_2 )x} \bar A_{12}(r_1,r_2)
		+e^{-i (r_1 k_1+r_2 k_2 )x} \bar A_{21}(r_1,r_2) 
	\right], 
\end{equation}
where we used~\eqref{eq:dagger} and~\eqref{eq:kredef}.  
Eq.~\eqref{eq:overlap} is valid for all the Bethe eigenstates~\eqref{eq:ansatz}. However, 
it is straightforward to check that the eigenstates with $\sigma=-1$ (see section~\eqref{sec:deph-ind}) 
have zero overlap with~\eqref{eq:initial}. 

Let us now restrict ourselves to the eigenstates forming the diffusive band (see 
Fig.~\ref{fig1:cartoon} (b)). As discussed in section~\ref{sec:diff}, 
the corresponding solutions of the Bethe equations form complex conjugated pairs, and 
can be treated by means of the string hypothesis. 
As anticipated in section~\ref{sec:norm}, an important issue is that upon 
substituting the solutions of the BGT equation~\eqref{eq:bgt-eq} in the 
expression for $|k_1,k_2\rangle$, $\langle k_1,k_2|$, 
spurious divergences appear. The divergences are due to the presence of the term 
$(\gamma/2-\sin(k_1)-\sin(k_2))^{-1}$. 
Moreover, both $N_1$ and $N_2$ in~\eqref{eq:norm1} and~\eqref{eq:norm2} diverge as well. 
To solve this issue, one can first exploit the invariance under reflections of $k_1,k_2$. 
Specifically, it is convenient to consider new quasimomenta 
\begin{equation}
	\label{eq:k-redef}
	k_1\to k_1,\quad k_2\to-k_2. 
\end{equation}
After using~\eqref{eq:k-redef}, one obtains that the term 
$(\gamma/2-\sin(k_1)+\sin(k_2))^{-1}$ is singular. This 
is convenient because now only  $N_1$ is singular, whereas $N_2$ (cf.~\eqref{eq:norm1}) is 
regular. To proceed, one has to determine the singularities of $|k_1,k_2\rangle$, $\langle k_1,k_2|$. 
The singularity structure  of the terms entering in $| k_1,k_2\rangle$ (cf.~\eqref{eq:ansatz}) 
is given as 
\begin{align}
	\label{eq:t1}
	 & e^{-ik_1x_1-ik_2x_2}A_{12}(-,-)\simeq 
	   \delta^{1-(x_1+x_2)/(2L)}\\
	   \label{eq:t2}
	& e^{-ik_1x_1+ik_2x_2}A_{12}(-,+)\simeq 
	 \delta^{-(x_1-x_2)/(2L)}\\
	\label{eq:t3}
	& e^{ik_1x_1-ik_2x_2}A_{12}(+,-)\simeq
	 \delta^{1+(x_1-x_2)/(2L)}\\
	 \label{eq:t4}
	& e^{ik_1x_1+ik_2x_2}A_{12}(+,+)\simeq	
	\delta^{(x_1+x_2)/(2L)}, 
\end{align}
where we assume $x_1\le x_2$, and we defined $\delta$ as 
\begin{equation}
	\label{eq:delta-def}
	\delta=\gamma/2-\sin(k_1)+\sin(k_2). 
\end{equation}
Notice that  $\delta={\mathcal O}(e^{-aL})$, with $a>0$. A similar result can be obtained for 
the terms with scattering amplitudes $A_{21}$ (cf.~\eqref{eq:ansatz}). 
Importantly, to derive~\eqref{eq:t1}~\eqref{eq:t2}~\eqref{eq:t3}~\eqref{eq:t4}, 
we employed the Bethe equations~\eqref{eq:beq-1} and~\eqref{eq:beq-2} to write the 
diverging contributions $e^{ik_j L}$ in terms 
of $\delta$. To proceed, we observe that a similar calculation can be done for the 
contributions appearing in $\langle k_1,k_2|$. Now, upon taking the limit  
$\delta\to0$ the singularities cancel out. Precisely, the term 
$N_1$ in the norm~\eqref{eq:norm1} diverges as $\delta^{-1}$, implying that  (cf.~\eqref{eq:G-exp})
$|k_1,k_2\rangle\langle k_1,k_2|0\rangle={\mathcal O}(\delta^{-1})$ for any $k_1,k_2$ satisfying 
the BGT equation~\eqref{eq:bgt-eq}.  A straightforward calculation shows that the only possibility is 
that $\langle k_1,k_2|0\rangle={\mathcal O}(\delta^{-1})$ and $|k_1,k_2\rangle={\mathcal O}(1)$. 
Precisely, only the term with $A_{12}(-,+)$ (cf.~\eqref{eq:ansatz}) survives in $|k_1,k_2\rangle$. 
Similarly, one has to keep only the terms with $\bar A_{12}(+,-)$ and $\bar A_{21}(+,-)$ in 
the overlap $\langle k_1,k_2|0\rangle$. After removing the singularities, 
the fermionic density $G_{x1,x1}$ is given as 
\begin{equation}
	\label{eq:density}
	G_{x_1,x_1}=\sum_{\{k_1,k_2\}}{\widetilde N_1}^{-1}e^{i(k_2- k_1) x_1}B_{12}(-,+)\bar g_{x,x}. 
\end{equation}
The sum in~\eqref{eq:density} is over the eigenstates of ${\mathcal L}^{\scriptscriptstyle(2)}$ forming 
the diffusive band (see section~\ref{sec:diff}), and which are treated within the framework of the 
string hypothesis. 
In~\eqref{eq:density} we defined 
\begin{equation}
	B_{12}(-,+)=\frac{e^{ik_1}(1-ie^{-ik_1}\gamma^-)}{e^{ik_1}-i\gamma^-}. 
\end{equation}
In~\eqref{eq:density} $\bar g_{x,x}$ is the finite part of the overlap with the initial condition, and it 
reads as 
\begin{equation}
	\bar g_{x,x}=e^{i(k_1-k_2)x}\bar B_{12}(+,-)+e^{-i(k_1-k_2)z}\bar B_{21}(+,-), 
\end{equation}
where 
\begin{align}
	&\bar B_{12}(+,-)=\frac{1-i\gamma^- e^{-i k_1}}{1-i \gamma^- e^{ik_1}}\frac{\gamma/2+\sin(k_1)-\sin(k_2)}{\gamma/2-\sin(k_1)-\sin(k_2)}
	(\gamma+2\sin(k_1)+2\sin(k_2))\\
	& \bar B_{21}(+,-)=-\frac{1+i\gamma^-e^{ik_2}}{1+i\gamma^- e^{-i k_2}}(\gamma+2\sin(k_1)-2\sin(k_2)). 
\end{align}
Finally, the finite part of the normalization ${\widetilde N}_1^{-1}$ in~\eqref{eq:density} is given as 
\begin{equation}
	{\widetilde N}_1^{-1}=
	\frac{4i\gamma(\cos(k_1)-\cos(k_2))(\gamma+\sin(k_1)+\sin(k_2))}{(\gamma/2+\sin(k_1)-\sin(k_2))(\gamma/2-\sin(k_1)-\sin(k_2))}L. 
\end{equation}
Again, Eq.~\eqref{eq:density} should hold in the long-time limit, provided that $L$ is large enough 
to ensure the validity of the string hypothesis.

\section{Numerical results}
\label{sec:numerics}

Here we provide numerical results supporting  the Bethe ansatz 
treatment of the previous sections. First, in section~\ref{sec:beq-num} 
we discuss the numerical solution of the Bethe equations~\eqref{eq:beq-1} 
and~\eqref{eq:beq-2}. In section~\ref{sec:bgt-eq} we focus on the 
Bethe-Gaudin-Takahashi equation~\eqref{eq:bgt} and~\eqref{eq:bgt-eq}. 
In section~\ref{sec:liouv} we discuss exact diagonalization data for  
the eigenvalues of ${\mathcal L}^{\scriptscriptstyle(2)}$. In section~\ref{sec:ed-vs-bethe} 
we compare the spectrum of ${\mathcal L}^{\scriptscriptstyle(2)}$ 
obtained from exact diagonalization with the Bethe ansatz results. 
In section~\ref{sec:gap} we investigate the finite-size scaling of the 
Liouvillian gap. 
In section~\ref{sec:num-G} we address the dynamics of the \emph{full} 
correlator $G_{x_1,x_2}$ (cf.~\eqref{eq:G}). In section~\ref{sec:profile-den} 
we focus on the profile of the fermion density. Finally, in section~\ref{sec:scaling} 
we discuss the diffusive scaling of the fermion density and its violation 
due to the boundary losses.

\subsection{Numerical solution of the Bethe equations}
\label{sec:beq-num}

The numerical solution of the Bethe equations is in general a challenging task. 
Indeed, Eq.~\eqref{eq:beq-1} and~\eqref{eq:beq-2} have $L^2$ solutions in the 
complex plane. Moreover, $k_1=0$ and $k_2=0$, as well as $k_1=\pi$ and $k_2=\pi$ 
are always solutions, although they have to be excluded because 
they correspond to vanishing eigenvectors. Similarly,  
the solutions with $k_1=k_2$ have to be excluded. Pairs of solutions $(k_1,k_2)$ 
that are related by a shift by $\pm\pi$ have to be counted only once. 
Crucially, since all the solutions of~\eqref{eq:beq-1} and~\eqref{eq:beq-2} 
are complex, multiple-precision arithmetic is necessary to evaluate the 
exponentials in the left-hand side of the equations. 
%
\begin{table}[ht]
\scriptsize
\centering
Solutions of Bethe equations for $L=2$ and $L=3$, with $\gamma=\gamma^-=1/10$\\[1ex]
\begin{tabular}{ccccc}
\toprule
$L$ & $\sigma$ & $k_1$ & $k_2$& $\varepsilon$  \\[0.3em]
\toprule
2& $+$ & $2.1919997294-0.1312558459i$   &  $1.1457120114-0.0351640458i$ &  $-0.25-1.9993749023i$   \\
2& $+$ & $1.1457120114+0.0351640458i$   &  $2.1919997294+0.1312558459i$ &  $-0.25+1.9993749023i$ \\
2& $+$ & $0.7866450552-0.0353040029i$   &  $0.7866450552+0.0353040029i$   &  $-0.2$ \\
2& $-$ & $2.0934365739-0.0576711461i$ &    $2.0934365739+0.0576711461i$ & $-0.3$ \\
\midrule
3 & $+$ & $1.28745845407-0.04147721049i$ &$1.28745845407+0.04147721049i$  & $-0.2593393435$ \\
3&  $+$ & $2.67275603750+0.00860942154i$ & $2.67275603750-0.00860942154i$  & $-0.1155608174$\\
3&  $+$ & $2.28508055892+0.03633954446i$ & $0.71654617450+0.08931362935i$  & $-0.162549919 -2.8251952277i$ \\
3&  $+$ & $0.71654617450-0.08931362935i$ & $2.28508055892-0.03633954446i$    & $-0.162549919 +2.8251952277i$ \\
3& $+$ & $2.36338604213-0.06913029032i$ &$1.57756565342+0.0014316741i$ &  $-0.2 -1.4142135623i$ \\
3& $+$ & $1.5775656534-0.00143167416i$ &$2.36338604213+0.0691302903i$ &  $-0.2 +1.4142135623i$ \\
3& $-$ & $0.7866450552-0.0353040029i$ & $0.78664505521+0.03530400298i$ &  $-0.2$ \\
3& $-$ & $1.5707963267-0.0499791900i$ & $2.35494759837+0.03530400298i$ &  $-0.25+1.4133294025i$ \\
3& $-$ & $2.3549475983-0.0353040029i$ & $1.57079632679+0.04997919006i$  & $-0.25-1.4133294025i$ \\
\bottomrule
\end{tabular}
\caption{Full set of solutions of the Bethe equations (cf.~\eqref{eq:beq-1}\eqref{eq:beq-2}) for with $L=2$ (first four rows) 
 and $L=3$ (last nine rows) sites, and with $\gamma=\gamma^-=1/10$. We show the two quasimomenta 
 $k_1,k_2$ and the associated energy $\varepsilon$.  The last solution for $L=2$ and the last 
 three solutions for $L=3$ correspond to $\sigma=-1$ 
 in~\eqref{eq:ansatz}, i.e., they are the same as for $\gamma=0$. Notice that by using the 
 symmetries of the Bethe equations, we fix $0<\mathrm{Re}(k_1)<\pi$ and $0<\mathrm{Re}(k_2)<\pi$. 
 Given a set of solutions $(k_1,k_2)$, one has that 
 $(k_2^*,k_1^*)$ is also a solution. Notice that for $\sigma=-1$ only $L(L-1)/2$ 
 solutions are allowed. 
}
\label{table:exact_bethe}
\end{table}
%

To illustrate the structure of the solutions of the Bethe equations, it is useful to 
focus on chains with small $L$. 
In Table~\ref{table:exact_bethe} we show the full set of solutions of~\eqref{eq:beq-1} and~\eqref{eq:beq-2} 
for $L=2$ and $L=3$. We only a consider fermionic chain with open boundary conditions and 
$\gamma=\gamma^-=1/10$. The second column of the table shows the eigenvalues of $\sigma$ 
(cf.~\eqref{eq:ansatz}). The states with $\sigma=-1$ are not affected by dephasing, as discussed in 
section~\ref{sec:deph-ind}. The number of solutions with $\sigma=-1$ is $L(L-1)/2$. 
The third and fourth column show the solutions $(k_1,k_2)$ of the Bethe equations. Importantly, the 
invariance of~\eqref{eq:beq-1} and~\eqref{eq:beq-2}  under the change of sign of $k_j$ and under shifts by $\pi$ 
can be used to fix $0<\mathrm{Re}(k_j)<\pi$. Notice also that the Bethe equations 
are not invariant under exchange $k_1\leftrightarrow k_2$. However, given a solution 
$(k_1,k_2)$, then $(k_2^*,k_1^*)$, with the star denoting complex conjugation, is also a 
solution. As it is clear from~\eqref{table:exact_bethe}, some of the solutions are 
formed by pairs of complex conjugated momenta $(k_1,k_1^*)$. These correspond to the real 
eigenvalues $\varepsilon$ of ${\mathcal L}^{\scriptscriptstyle(2)}$. 
The last column in Table~\ref{table:exact_bethe} 
is the energy $\varepsilon$ as obtained by using~\eqref{eq:eps}. We checked that 
the eigenvalues coincide with the exact diagonalization results to machine precision. 

%
\begin{figure}[t]
\centering
\includegraphics[width=.85\linewidth]{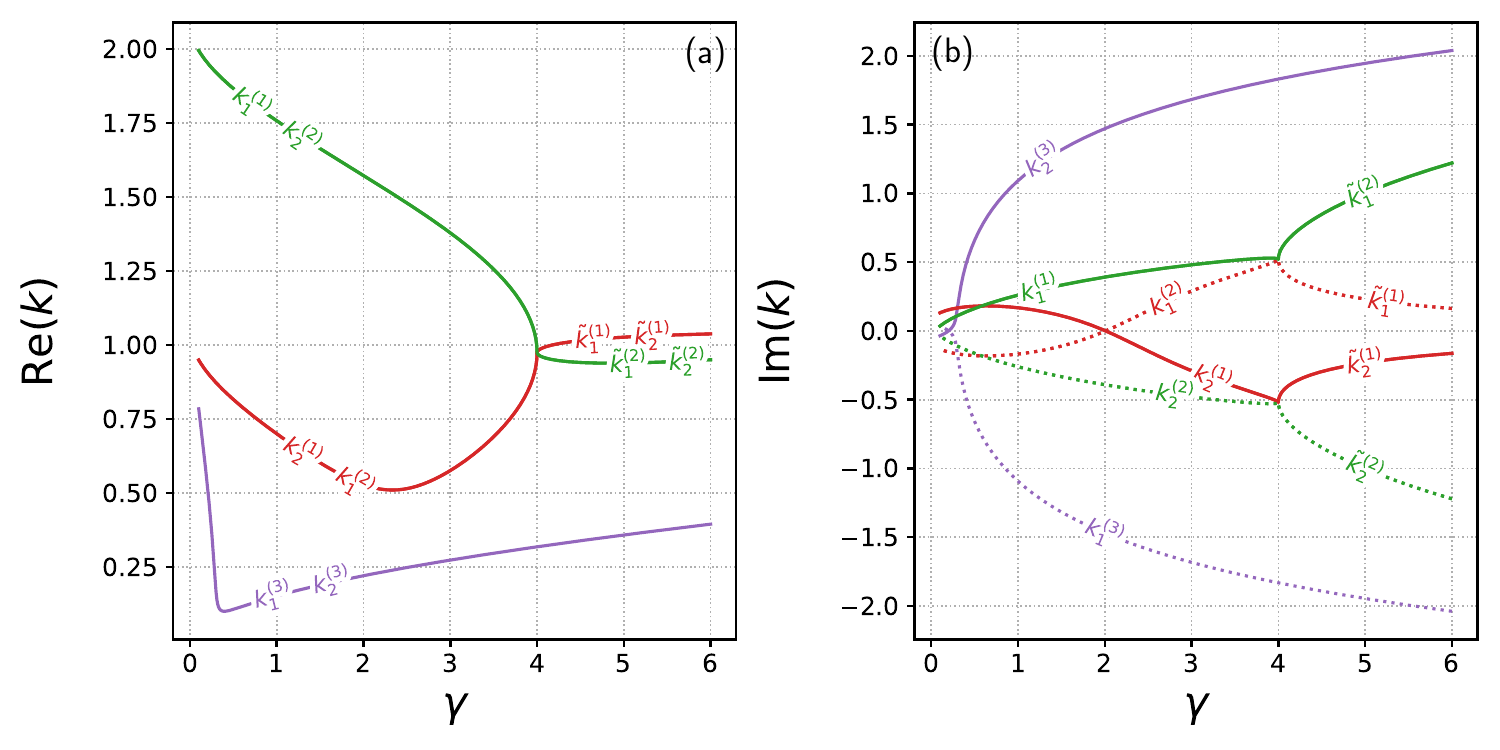}
\caption{ Solutions $(k_1,k_2)$ of the Bethe equations~\eqref{eq:beq-1} and~\eqref{eq:beq-2} 
 for $L=2$ and fixed $\gamma^-=1/10$. We only consider solutions  with 
 $\sigma=1$ (cf.~\eqref{eq:ansatz}). We plot $(k_1,k_2)$ as  functions of the 
 dephasing rate $\gamma$. Panel $(a)$ shows the real parts of $k_1$ and $k_2$, the imaginary parts 
 being reported in panel $(b)$. Starting from $\gamma=0$ one has the three pairs 
 $(k_1^{\scriptscriptstyle(j)},k_2^{\scriptscriptstyle(j)})$ with 
 $j=1,2,3$. The solutions with $j=1,2$ are related 
 by $(k^{\scriptscriptstyle(1)}_1,k_2^{\scriptscriptstyle(1)})=(k_2^{\scriptscriptstyle(2)},
 k_1^{\scriptscriptstyle(2)})^*$, with the star denoting complex conjugation. 
 The solution with $j=3$ is such that $k_1^{\scriptscriptstyle(3)}=(k_2^{\scriptscriptstyle(3)})^*$. At 
 $\gamma=4$ the solutions with $j=1,2$ ``collide''. The solutions at $\gamma>4$ are denoted with 
 a tilde. At $\gamma>4$ all the solutions are formed by complex conjugate pairs, i.e., 
 $\tilde k_1^{\scriptscriptstyle(j)}=(\tilde k^{\scriptscriptstyle(j)}_2)^*$ 
 for any $j$. 
}
\label{fig:beq-L2}
\end{figure} 
%

It is interesting to investigate how the solutions of the Bethe equations 
change as a function of dissipation. In Fig.~\ref{fig:beq-L2} we show the 
solutions of the Bethe equations for $L=2$ at fixed $\gamma^-=1/10$ as a function of $\gamma$. 
We only consider the three solutions with $\sigma=1$ (cf.~\eqref{eq:ansatz} and Table~\ref{table:exact_bethe}). 
Panel $(a)$ and $(b)$ show $\mathrm{Re}(k_j)$ and $\mathrm{Im}(k_j)$ as a function 
of $\gamma$. We consider the interval $\gamma\in [1/10,6]$. We denote the 
different solutions as $(k_1^{\scriptscriptstyle (p)},k_2^{\scriptscriptstyle (p)})$, 
with $p\in [1,3]$. 
Now, the solution at the bottom in Fig.~\ref{fig:beq-L2} (a), i.e., with $p=3$, 
corresponds to $k_2=k_1^*$, i.e., to  real energy $\varepsilon$. 
The remaining two solutions are such that 
$k^{\scriptscriptstyle(1)}_1=(k_2^{\scriptscriptstyle(2)})^*$ and 
$k^{\scriptscriptstyle(2)}_1=(k_2^{\scriptscriptstyle(1)})^*$. Interestingly, 
the behavior of the solutions  as a function of $\gamma$ is not ``smooth''. Precisely, 
at $\gamma=4$ the solutions with $p=1$ and $p=2$ ``collide'', whereas the one with 
$p=3$ remains isolated. At $\gamma>4$ the solutions with $p=1,2$ get reorganized. 
Precisely, they emerge as new pairs of solutions  $(\tilde k^{\scriptscriptstyle(p)}_1,\tilde 
k_2^{\scriptscriptstyle(p)})$, with $p=1,2$ for $\gamma>4$. Notice that for  
$\gamma>4$ all the three solutions are formed by 
complex conjugated momenta, i.e.,  $\tilde k_1^{\scriptscriptstyle (p)}=
(\tilde k_2^{\scriptscriptstyle(p)})^*$ for any $p$.

\subsection{Numerical solution of the Bethe-Gaudin-Takahashi (BGT) equation}
\label{sec:bgt-eq}

%
\begin{figure}[t]
\centering
\includegraphics[width=.6\linewidth]{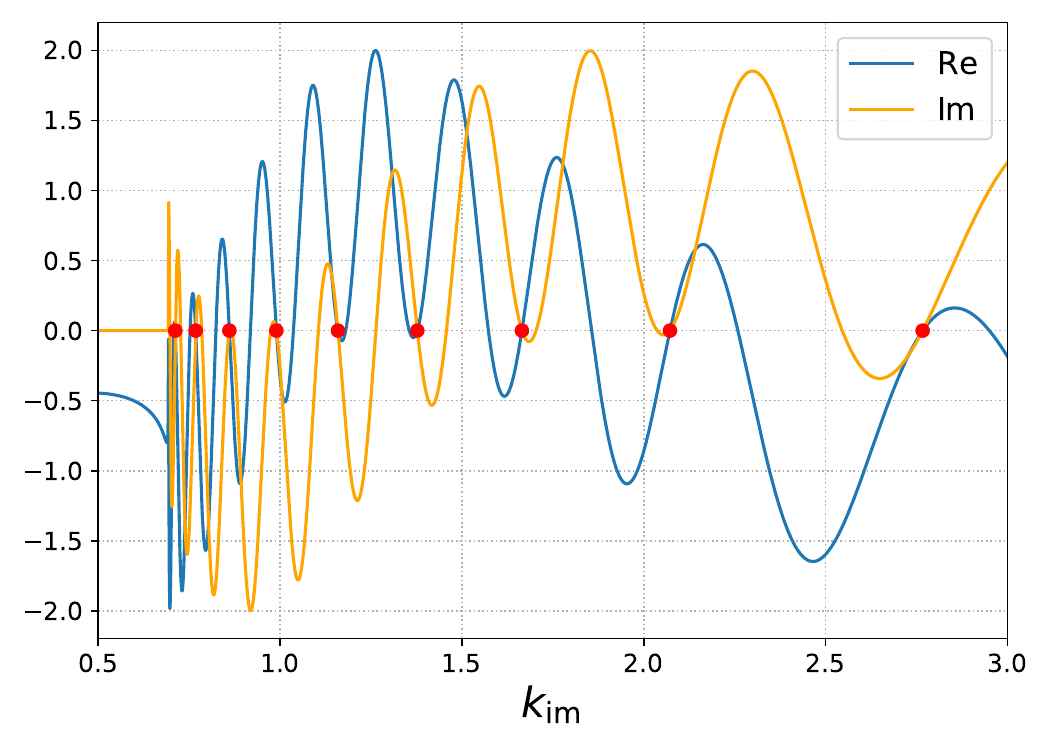}
\caption{ Numerical solution of the BGT equation~\eqref{eq:bgt} for 
 a fermionic chain with $L=10$  and $\gamma=5$ and $\gamma^-=0$. 
 The two curves show the real and imaginary parts of~\eqref{eq:bgt}, 
 plotted versus the imaginary part $k_\mathrm{im}$ of the quasimomenta 
 $k_1,k_2$. The circles are the solutions of the BGT equations. Notice that 
 there are $L-1$ solutions. The missing solution gives $\varepsilon=0$, and it  
 corresponds to $k_\mathrm{im}\to\infty$. 
}
\label{fig:bethe-num}
\end{figure} 
%

%
\begin{figure}[t]
\centering
\includegraphics[width=.6\linewidth]{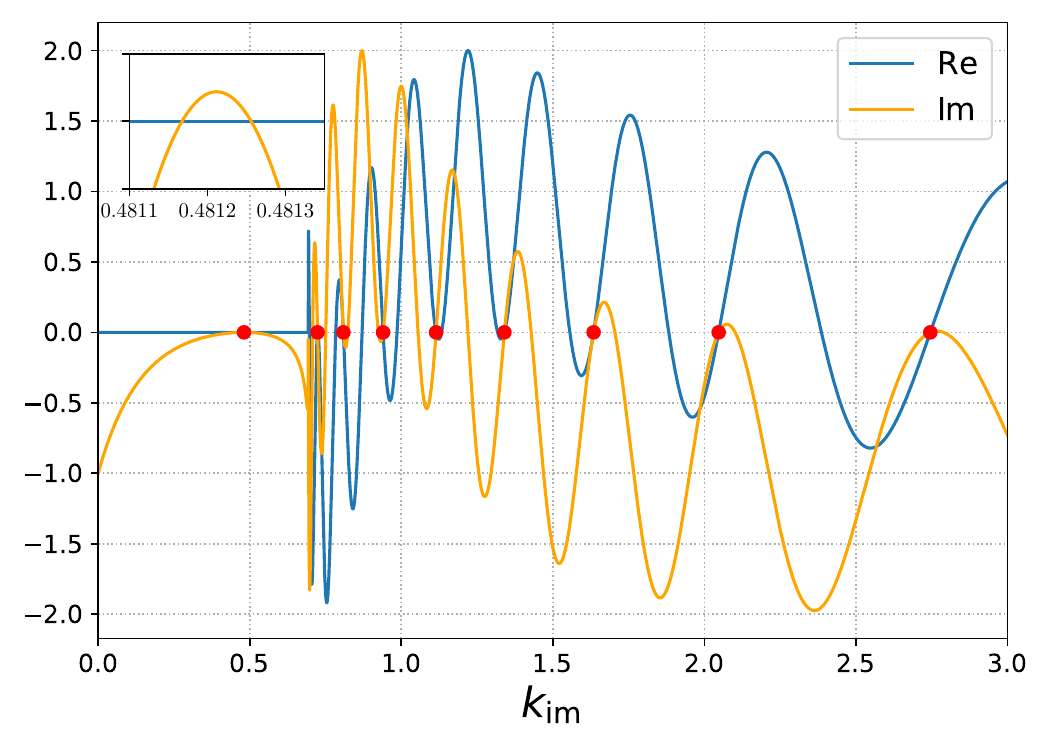}
\caption{ Same as in Fig.~\ref{fig:bethe-num} for $\gamma^-=1$. 
 Now there are $L-2$ solutions within the diffusive band. The 
 solution with $\varepsilon=0$ is not present for nonzero $\gamma^-$. 
 The two almost degenerate solutions at $k_\mathrm{im}\approx 0.5$ (see inset in the 
 Figure) correspond to the boundary-related eigenvalues of the Liouvillian 
 (star symbol in Fig.~\ref{fig1:cartoon} (b)). 
}
\label{fig:bethe-num-1}
\end{figure} 
%

The numerical results of the previous section showed that extracting the full set of 
solutions of the Bethe equations can be a challenging task, as expected. 
Here we focus on the solutions of the Bethe equations forming the diffusive band (see Fig.~\ref{fig1:cartoon} and 
section~\ref{sec:diff}). These solutions dominate the long-time behavior 
of physical observables, such as the fermion correlator $G_{x_1,x_2}$. 
In the limit of large $L$, one can use the string hypothesis~\cite{takahashi1999thermodynamics}. Thus, 
the solutions of the Bethe equations forming the diffusive band are well approximated by the solutions 
of the Bethe-Gaudin-Takahashi equation~\eqref{eq:bgt}. Solving~\eqref{eq:bgt} is a much simpler 
task because Eq.~\eqref{eq:bgt} is only function of $k_\mathrm{im}$, which is 
real. Importantly, by using the BGT equation in logarithmic form~\eqref{eq:bgt-eq} and by varying the 
quantum numbers $I_j$, one can target 
the different momentum pairs $(k_1,k_2)$ forming the diffusive band. 

Here we focus on the numerical solution of the BGT equation~\eqref{eq:bgt}. In Fig.~\ref{fig:bethe-num} we plot the real and imaginary 
parts (curves with different colors) of~\eqref{eq:bgt} for a chain with $L=10$, $\gamma=5$ and $\gamma^-=0$. 
On the $x$-axis $k_\mathrm{im}$ is the imaginary part of $k_1=k_2^*$. The real part is obtained 
from~\eqref{eq:string}. 
The simultaneous crossing (full circles) of the two curves with the horizontal axis marks the solutions of the 
BGT equation. Notice that there are $L-1$ solutions. The missing solution is that with ${\varepsilon=0}$. 
This is present only for $\gamma^-=0$ and it corresponds to diverging $k_1,k_2$. 
Let us now investigate the effect of the losses. 
In Fig.~\ref{fig:bethe-num-1} we show the numerical solutions 
of~\eqref{eq:bgt} for $L=10$, $\gamma=5$ and $\gamma^-=1$. Now, there are $L$ solutions. The solution 
with $\varepsilon=0$ is not present. 
Interestingly, the leftmost circle in Fig.~\ref{fig:bethe-num-1} corresponds to two 
almost degenerate solutions of~\eqref{eq:bgt}. These are the boundary-related eigenvalues 
of ${\mathcal L}^{\scriptscriptstyle(2)}$ discussed in section~\ref{sec:diff}.  
These boundary states are present only for $\gamma^- >\exp({-\mathrm{arccosh(\gamma/4)}})$. 
Upon lowering $\gamma^-$ they  merge with the diffusive band (cf.~\ref{fig1:cartoon}). 
The inset of Fig.~\ref{fig:bethe-num-1} shows a zoom of the real and imaginary 
parts of the BGT equation~\eqref{eq:bgt} 
around the two degenerate solutions (leftmost circle in the main Figure). For 
$L=10$ the difference between the two solutions is ${\mathcal O}(10^{-4})$.

\subsection{Spectrum of the Liouvillian ${\mathcal L}^{\scriptscriptstyle(2)}$: Overview}
\label{sec:liouv}

%
\begin{figure}[t]
\centering
\includegraphics[width=1\linewidth]{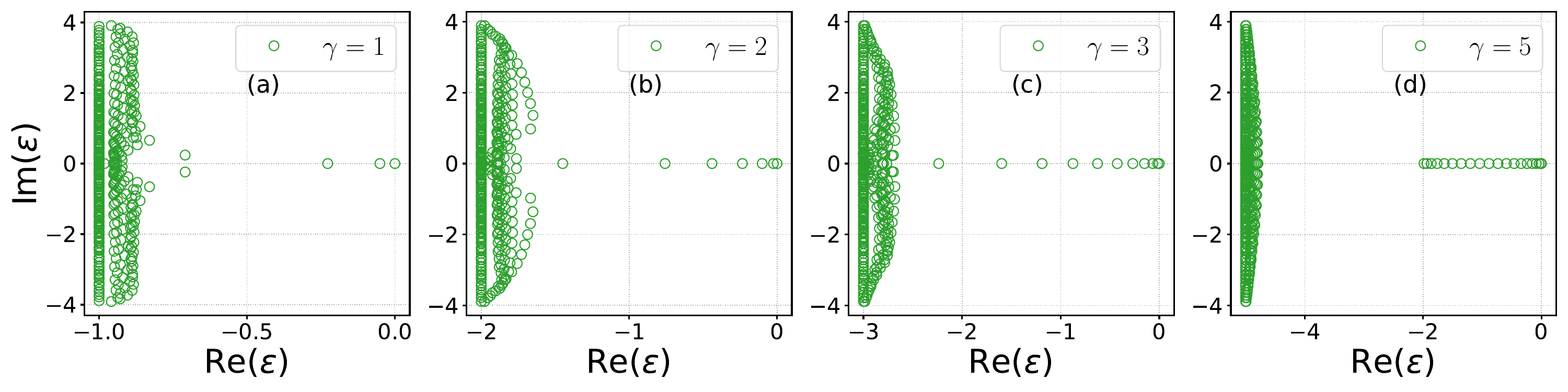}
\caption{ Spectrum of the Liouvillian ${\mathcal L}^{\scriptscriptstyle(2)}$ for a 
 fermionic chain with bulk dephasing and no boundary losses, i.e., $\gamma^-=0$. 
 The symbols are exact diagonalization results for $L=20$. 
 We plot $\mathrm{Im}(\varepsilon)$ 
 versus ${\mathrm{Re}(\varepsilon)}$. The different panels show 
 different values of $\gamma$. As one increases $\gamma$ (from left to right 
 in the Figure), the diffusive band with real eigenvalues gets populated and 
 well separated from the rest of the spectrum. 
}
\label{fig:panel}
\end{figure} 
%

Here we illustrate the general structure of the spectrum of the Liouvillian ${\mathcal L}^{\scriptscriptstyle(2)}$ 
for a fermionic chain with open boundary conditions with $\gamma^-=0$ as a function of $\gamma$. 
In Fig.~\ref{fig:panel} we report exact diagonalization results for a chain with $L=20$ and no boundary losses, i.e., 
with $\gamma^-=0$. 
First, for $\gamma=\gamma^-=0$ the $L^2$ eigenvalues 
$\varepsilon$ form a straight line parallel to the imaginary axis (not shown in the Figure). Indeed, for $\gamma=0$,  $k_1,k_2$ are solutions of~\eqref{eq:beq-3} 
and~\eqref{eq:beq-4} with $\gamma^-=0$, and are real. As a consequence the 
eigenvalues $\varepsilon$ (cf.~\eqref{eq:eps}) have the same real part $-\gamma$. As discussed in section~\ref{sec:deph-ind}, at finite $\gamma$ there are $L(L-1)/2$ momentum pairs $(k_1,k_2)$ that 
remain the same as for $\gamma=0$. This implies that the eigenvalues $\varepsilon$ are the same as for $\gamma=0$, 
apart from a trivial shift by $-\gamma$ (cf.~\eqref{eq:eps}). These eigenvalues correspond to the vertical straight 
lines with $\mathrm{Re}(\varepsilon)=-\gamma$ in the different panels. Near this vertical lines there 
are $\sim L(L-1)/2$ eigenvalues that depend in a nontrivial way on $\gamma$. 
These eigenvalues correspond to complex solutions of the Bethe equations~\eqref{eq:beq-1} and~\eqref{eq:beq-2} 
with vanishing imaginary parts in the limit $L\to\infty$. For large $L$ these eigenvalues can be understood 
perturbatively in $1/L$, as it was discussed in section~\ref{sec:vanishing}, at least for large enough $\gamma$. 
Finally, upon increasing $\gamma$ a band of 
real eigenvalues appear. This band contains the eigenvalues that are responsible for the diffusive spreading of 
particles at long times. A similar band appears in the tight-binding chain with periodic boundary conditions~\cite{esposito2005emergence}, 
and for the periodic tight-binding chain subject to incoherent hopping~\cite{eisler2011crossover}. 
As $\gamma$ increases, the separation in energy between  the diffusive 
band and the remaining part of the spectrum increases. Similar separation in different connected components of the Liouvillian 
spectrum has been observed in random Liouvillians~\cite{costa2022spectral}. 
As discussed in section~\ref{sec:diff}, at $\gamma>4$ the diffusive band contains 
at least $L-2$ states, whereas the number of states diminishes upon lowering $\gamma$. These eigenvalues in the 
diffusive band correspond to string solutions of the Bethe equations~\eqref{eq:beq-1} and~\eqref{eq:beq-2}, and can be 
effectively treated within the framework of the string hypothesis (see section~\ref{sec:diff}).

\subsection{Spectrum of ${\mathcal L}^{\scriptscriptstyle(2)}$: Exact diagonalization versus Bethe ansatz}
\label{sec:ed-vs-bethe}

%
\begin{figure}[t]
\centering
\includegraphics[width=.49\textwidth]{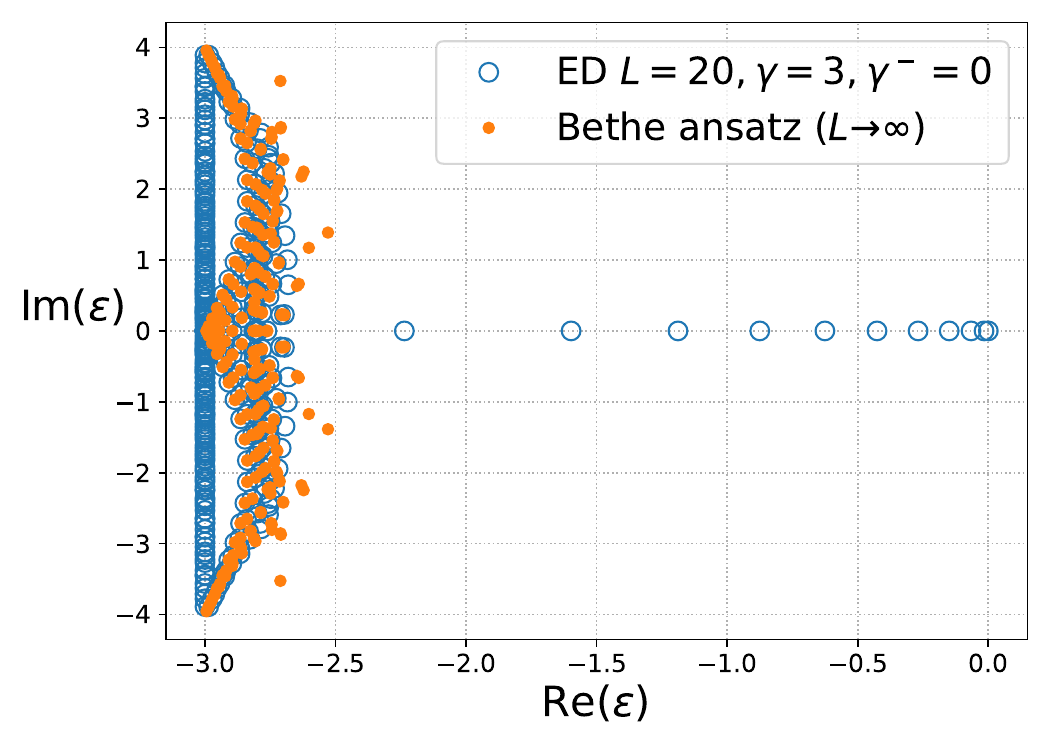}
\includegraphics[width=.49\textwidth]{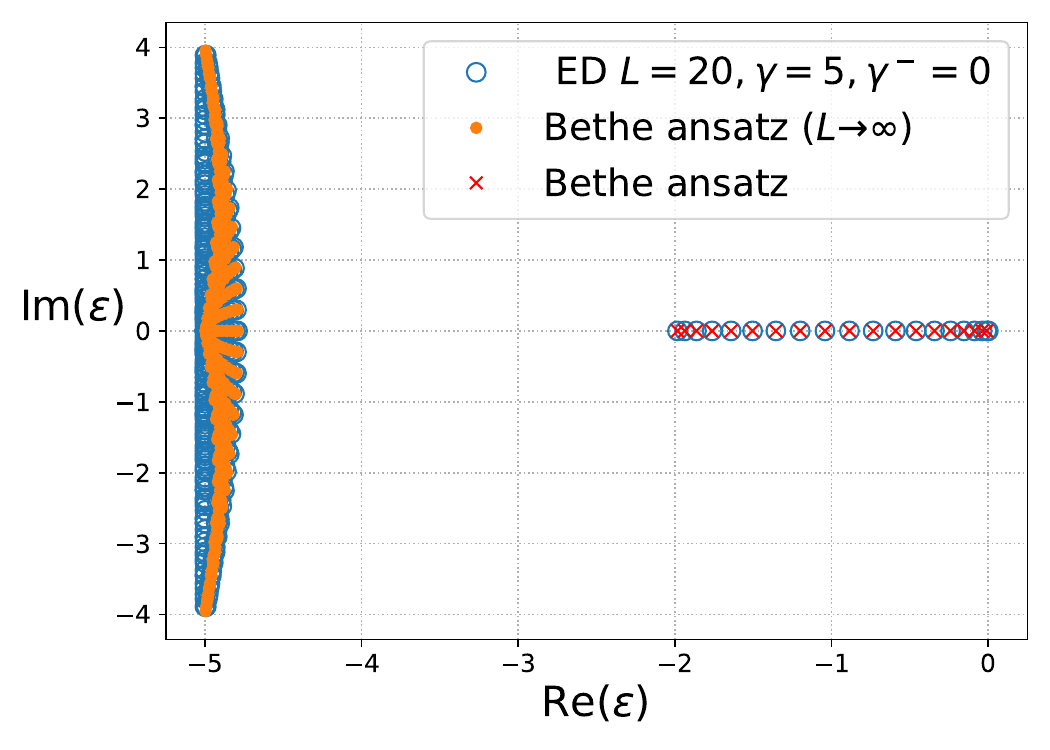}
\caption{ Spectrum of the Liouvillian ${\mathcal L}^{(2)}$ for the open fermionic chain 
 with $L=20$ and $\gamma^-=0$. The left (right) panel shows exact diagonalization results (circles) 
 for $\gamma=3$ ($\gamma=5$).  In both panels a vertical band of $L(L-1)/2$ eigenvalues is present 
 at $\varepsilon=-\gamma$. These eigenvalues are the same as for $\gamma^-=0$, apart from a trivial 
 shift by $-\gamma$. Near the vertical band a connected region with $\sim L^2/2$ eigenvalues is present. 
 This correspond to the complex solutions of the Bethe equations with vanishing imaginary parts 
 (see section~\ref{sec:vanishing}). Finally, a diffusive band of real eigenvalues is also visible in 
 both panels. For $\gamma=5$  (right panel) the diffusive band is well separated from the rest of the spectrum, and it 
 contains $L$ states.  
 The large $L$ expansion~\eqref{eq:k1-large-L} and~\eqref{eq:k2-large-L} is reported in both panels with 
 the full circles.  The expansion describes well the eigenvalues near the vertical band. 
 The crosses in the right panel are Bethe ansatz results 
 obtained by solving numerically~\eqref{eq:baer} and~\eqref{eq:baei} using the solutions of the 
 BGT equation~\eqref{eq:bgt-eq} as initial guess. 
}
\label{fig:EDL20g5}
\end{figure} 
%

%
\begin{figure}[t]
\centering
\includegraphics[width=.7\linewidth]{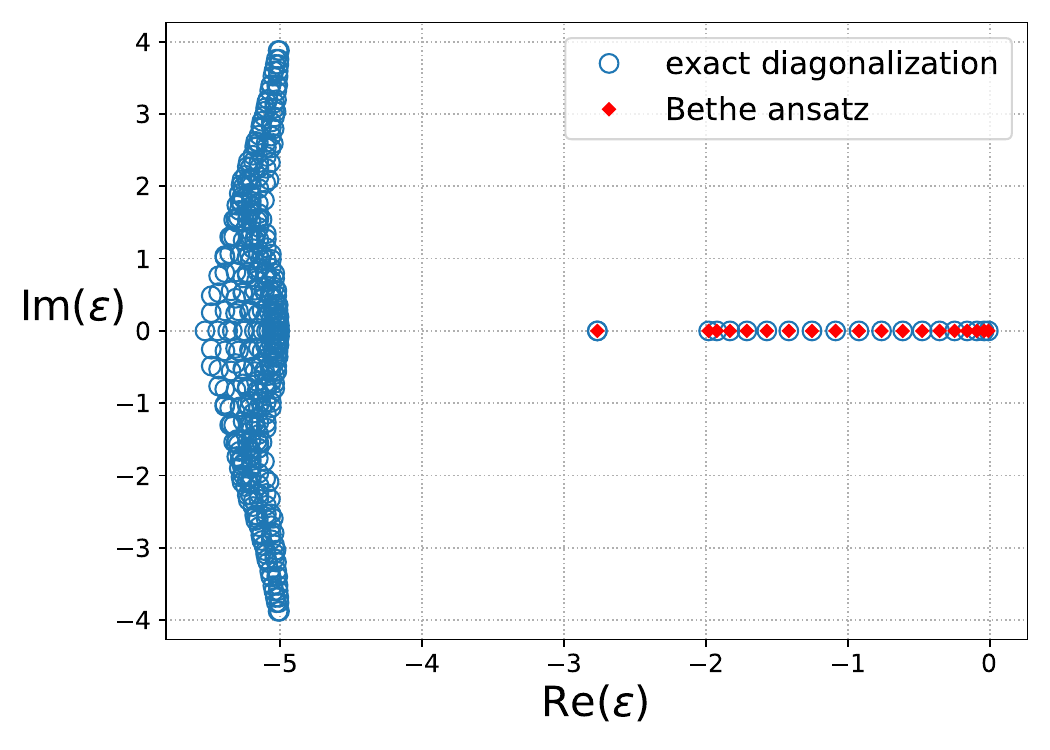}
\caption{ Same as in Fig.~\ref{fig:EDL20g5} for $L=20$, and  $\gamma=5$ and 
 $\gamma^-=1$. The diffusive band with $\varepsilon>\sqrt{\gamma^2-16}-\gamma$ contains 
 $L-2$ solutions. An isolated doublet of quasi-degenerate eigenvalues is 
 present at $\varepsilon\approx -2.7$, and it corresponds to the boundary-related eigenvalues 
 in Fig.~\ref{fig1:cartoon} (b) (see full diamond  symbol in the Figure). 
 The empty circles in the Figure are exact diagonalization (ED) results. 
 The diamonds are the results obtained by solving the BGT 
 equation~\eqref{eq:bgt} numerically. 
}
\label{fig:EDL20gp}
\end{figure} 
%

Let us now compare the Bethe ansatz results for the eigenvalues $\varepsilon$ of the Liouvillian ${\mathcal L}^{\scriptscriptstyle (2)}$ 
and exact diagonalization (ED) data. In Fig.~\ref{fig:EDL20g5} we show ED data for $L=20$, $\gamma^-=0$ and 
$\gamma=3$ and $\gamma=5$ (left and right panel, respectively). In both panels there is a vertical 
band containing $L(L-1)/2$ eigenvalues. These are the same, except for a trivial shift by $-\gamma$, 
as for the open fermionic chain with $\gamma=0$.  
At small $\mathrm{Re}(\varepsilon)$ the  diffusive band of real eigenvalues is visible. 
The complex eigenvalues $\varepsilon$ between the diffusive 
band and the vertical band correspond to complex solutions of the Bethe equations with 
vanishing imaginary parts in the limit $L\to\infty$. As discussed in section~\ref{sec:vanishing} 
these eigenvalues of ${\mathcal L}^{\scriptscriptstyle (2)}$ can be understood perturbatively in 
$1/L$. 
The full circles in the Figure are  obtained from the large $L$ expansions~\eqref{eq:k1-large-L} and~\eqref{eq:k2-large-L}. 
Fig.~\ref{fig:EDL20g5} shows that the large $L$ expansion works well at $\gamma=5$, i. e., 
when the diffusive band is well separated from the rest of the spectrum.  However, 
when the diffusive band overlaps with the other regions of the spectrum, the agreement 
between the large $L$ expansions and the ED data is not perfect. 

For $\gamma=5$ in Fig.~\ref{fig:EDL20g5} we report with the crosses 
the Bethe ansatz results for the eigenvalues of the diffusive band.
As it was stressed in section~\ref{sec:beq-num}, extracting the full set of solutions of 
the Bethe equations is a challenging task. A convenient strategy for the diffusive band is to 
first solve the BGT equation~\eqref{eq:bgt}, and then use the solutions as initial guess to 
solve~\eqref{eq:beq-1} and~\eqref{eq:beq-2}. 
The agreement between the Bethe ansatz results and the ED data is perfect. A similar agreement is found for $\gamma=3$, 
although we do not report the results in the Figure. 
Finally, let us discuss the effect of $\gamma^-$. In Fig.~\ref{fig:EDL20gp} we show 
ED results for $L=20$, $\gamma=5$ and $\gamma^-=1$. Now, the 
vertical band at small eigenvalues $\varepsilon\approx -\gamma$ is deformed. Still, there are $L(L-1)/2$ 
eigenvalues that are the same as in the case with $\gamma=0$, apart from the trivial shift by $-\gamma$. 
Again, a diffusive band is present at large  $\mathrm{Re}(\varepsilon)\approx 0$. The band contains $L$ energies. 
The full diamonds are the results obtained using the string hypothesis, 
i.e., by solving the BGT equation~\eqref{eq:bgt}. The agreement with the ED data is 
perfect, even at finite $\gamma^-$ and despite the fact that the BGT equation~\eqref{eq:bgt} is 
valid only in the limit $L\to\infty$. Finally, the real energy at $\varepsilon\approx -2.7$ 
corresponds to the almost degenerate doublet of boundary-related eigenvalues of the Liouvillian 
(see section~\ref{sec:diff}), which appear for $\gamma^->\exp({-\mathrm{arccosh}(\gamma/4)})$. 

\subsection{Finite-size scaling of the Liouvillian gap}
\label{sec:gap}

%
\begin{figure}[t]
\centering
\includegraphics[width=.7\linewidth]{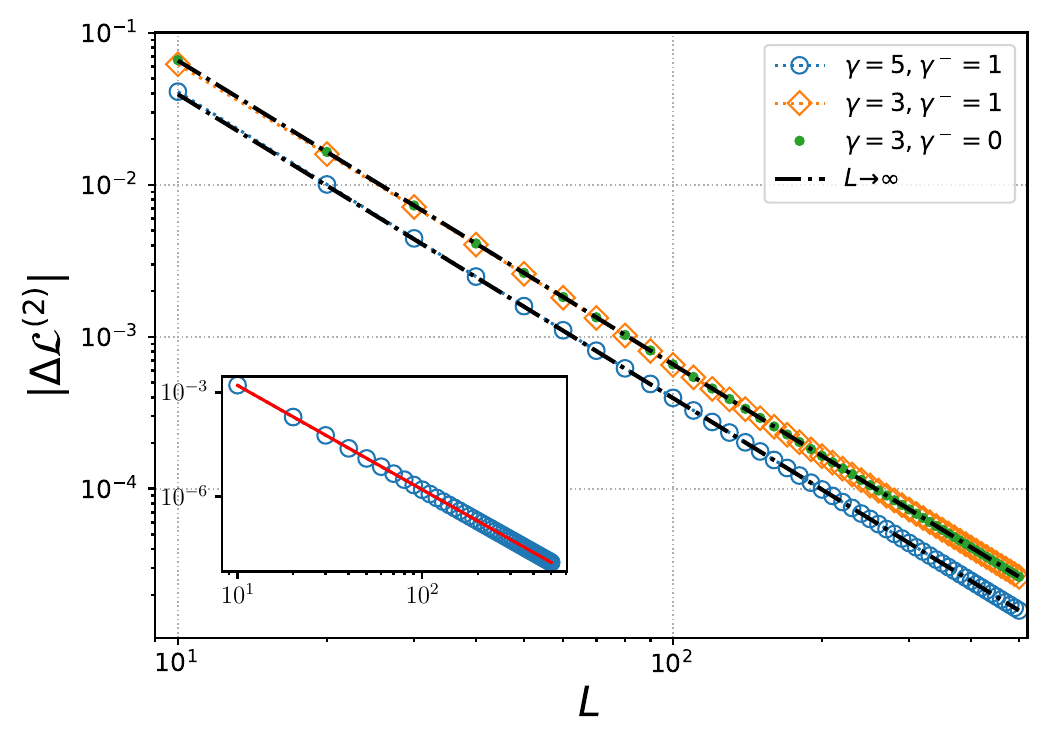}
\caption{ Finite-size scaling of the Liouvillian gap defined in~\eqref{eq:gap-2}. 
 We show results for several values of the dephasing 
 rate $\gamma$ and boundary loss rate $\gamma^-$ (different symbols in the 
 Figure). We consider chains with $L\lesssim 500$. The dashed-dotted line is 
 the Bethe ansatz prediction~\eqref{eq:liou-gap} in the limit $L\to\infty$. 
 In the inset we focus on the subleading contributions to the gap for the 
 case with $\gamma=5$ and $\gamma^-=1$. The symbols are $\Delta{\mathcal L}^{\scriptscriptstyle(2)}+2\pi^2/(\gamma L^2)$ 
 plotted versus $L$. The continuous line is the analytic prediction (cf. second term in~\eqref{eq:liou-gap}). 
}
\label{fig:gap}
\end{figure} 
%

Let us now discuss the finite-size scaling of the gap of the Liouvillian ${\mathcal L}^{\scriptscriptstyle (2)}$. 
The Liouvillian gap $\Delta{\mathcal L}^{\scriptscriptstyle(2)}$ is the eigenvalue of ${\mathcal L}^{\scriptscriptstyle(2)}$ 
with the largest \emph{nonzero} real part, i.e., 
\begin{equation}
	\label{eq:gap-2}
	\Delta{\mathcal L}^{\scriptscriptstyle(2)}:= -\max_{j}\mathrm{Re}(\varepsilon_j),\quad\mathrm{with}\,\,
	\mathrm{Re}(\varepsilon_j)\ne0. 
\end{equation}
As it is clear from Fig.~\ref{fig:EDL20g5} and Fig.~\ref{fig:EDL20gp}, the gap coincides with the 
largest nonzero energy in the diffusive band. 
For $\gamma^-=0$ and in the large $L$ limit the gap  is obtained by 
solving the BGT equation~\eqref{eq:bgt-eq} with $I_j=1$. For $\gamma^->0$ one has to fix $I_j=0$ in~\eqref{eq:bgt-eq}. 
In Fig.~\ref{fig:gap} we plot the Liouvillian gap as a function 
of $L$. We show results for $\gamma=3,5$ and $\gamma^-=0,1$. The symbols are the exact numerical 
data obtained from the BGT equation~\eqref{eq:bgt-eq}. 
The dashed-dotted lines are the results~\eqref{eq:liou-gap} in the large $L$ limit. 
The leading behavior as $\propto 1/L^2$ is visible. Notice that although Eq.~\eqref{eq:liou-gap} was 
derived for the case with nonzero $\gamma^-$ (see section~\ref{sec:diff}), it works also for $\gamma^-=0$. 
In the inset in Fig.~\ref{fig:gap} we focus on subleading terms, subtracting from 
$\Delta{\mathcal L}^{\scriptscriptstyle (2)}$ the leading $1/L^2$ behavior (cf.~\eqref{eq:liou-gap}).
Precisely, we plot $\Delta{\mathcal L}^{\scriptscriptstyle(2)}+2\pi^2/(\gamma L^2)$ versus $L$. We only consider the case 
with $\gamma=5$ and $\gamma^-=1$. The continuous line is the second term in~\eqref{eq:liou-gap}, which perfectly 
matches the data.

\subsection{Dynamics of the fermionic correlator: Bethe ansatz versus exact diagonalization}
\label{sec:num-G}

Having compared Bethe ansatz versus exact diagonalization results for the spectrum of the 
Liouvillian, we now focus on the time-dependent fermionic correlator $G_{x_1,x_2}(t)$. 
%
\begin{figure}[t]
\centering
\includegraphics[width=.7\linewidth]{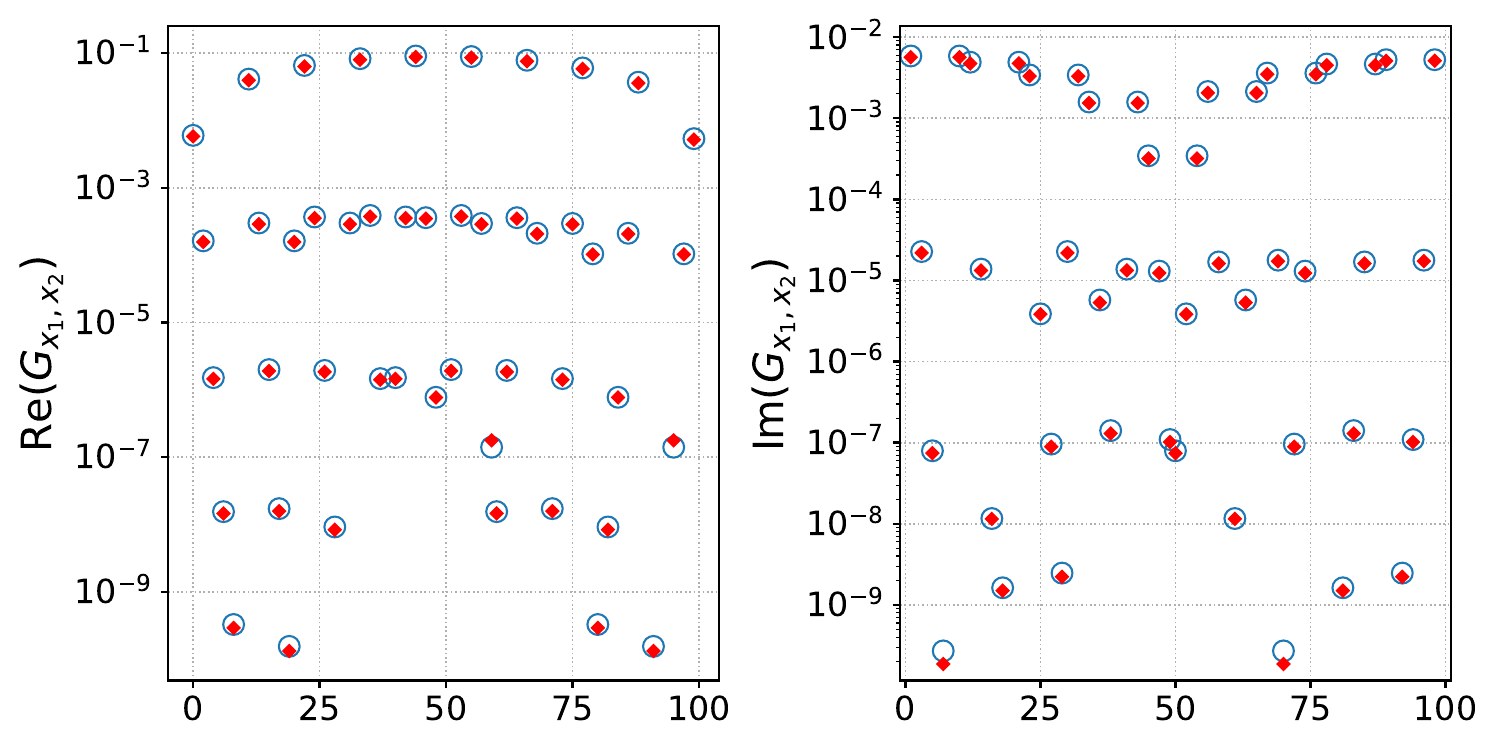}
\caption{ Fermionic correlator $G_{x_1,x_2}$ as a function of 
 time $t$. The left and right panels show the real and imaginary 
 parts of the correlator, respectively. The initial condition 
 is $G_{x_1,x_2}=\delta_{x_1,L/2}\delta_{x_2,L/2}$. The empty 
 circles are exact diagonalization (ED) results for the 
 matrix elements of  $G_{x_1,x_2}$ at $t=20$. Results are for a chain 
 with $L=10$, $\gamma=5$ and $\gamma^-=1$. The full 
 diamonds are the Bethe ansatz results. These are obtained 
 by solving the Bethe equations~\eqref{eq:beq-1} and~\eqref{eq:beq-2} 
 and using~\eqref{eq:G-exp}, where the sum is restricted to the 
 eigenvalues of the diffusive band. 
}
\label{fig:full-g}
\end{figure} 
%
Here we compare ED data versus Bethe ansatz results for the \emph{full} correlator $G_{x_1,x_2}$. 
The circles in Fig.~\ref{fig:full-g} are ED data for a chain with $L=10$ 
sites, $\gamma=5$, and $\gamma^-=1$. The results are at fixed time $t=20$. The left and right panels 
in the Figure show $\mathrm{Re}(G_{x_1,x_2})$ and $\mathrm{Im}(G_{x_1,x_2})$, respectively. Notice that 
on the $y$-axis we employ a logarithmic scale. The numbers on the real axis label the different 
matrix elements of $G_{x_1,x_2}$. The full diamonds are Bethe ansatz results. Precisely, 
here we obtain the full-time dynamics of $G_{x_1,x_2}$ by using the expansion~\eqref{eq:G-exp}. 
However, since finding all the solutions of the Bethe equations~\eqref{eq:beq-1} and~\eqref{eq:beq-2} 
is a daunting task, we truncate~\eqref{eq:G-exp} restricting the sum over the eigenvalues  in 
the diffusive band, which is expected to dominate the long-time behavior of the correlator. 
We first numerically solve the BGT equation~\eqref{eq:bgt-eq}, using the solutions as initial guess 
for the exact Bethe equations~\eqref{eq:beq-1} and~\eqref{eq:beq-2}. As it is clear from Fig.~\ref{fig:full-g}, 
the agreement between the Bethe ansatz and the exact diagonalization data is quite satisfactory. 
Deviations are present for the smaller matrix elements, and can be attributed to the complex eigenvalues 
$\varepsilon$ of the Liouvillian, which we are neglecting. 

%
\subsection{Dynamics of the density profile}
\label{sec:profile-den}

%
\begin{figure}[t]
\centering
\includegraphics[width=.95\linewidth]{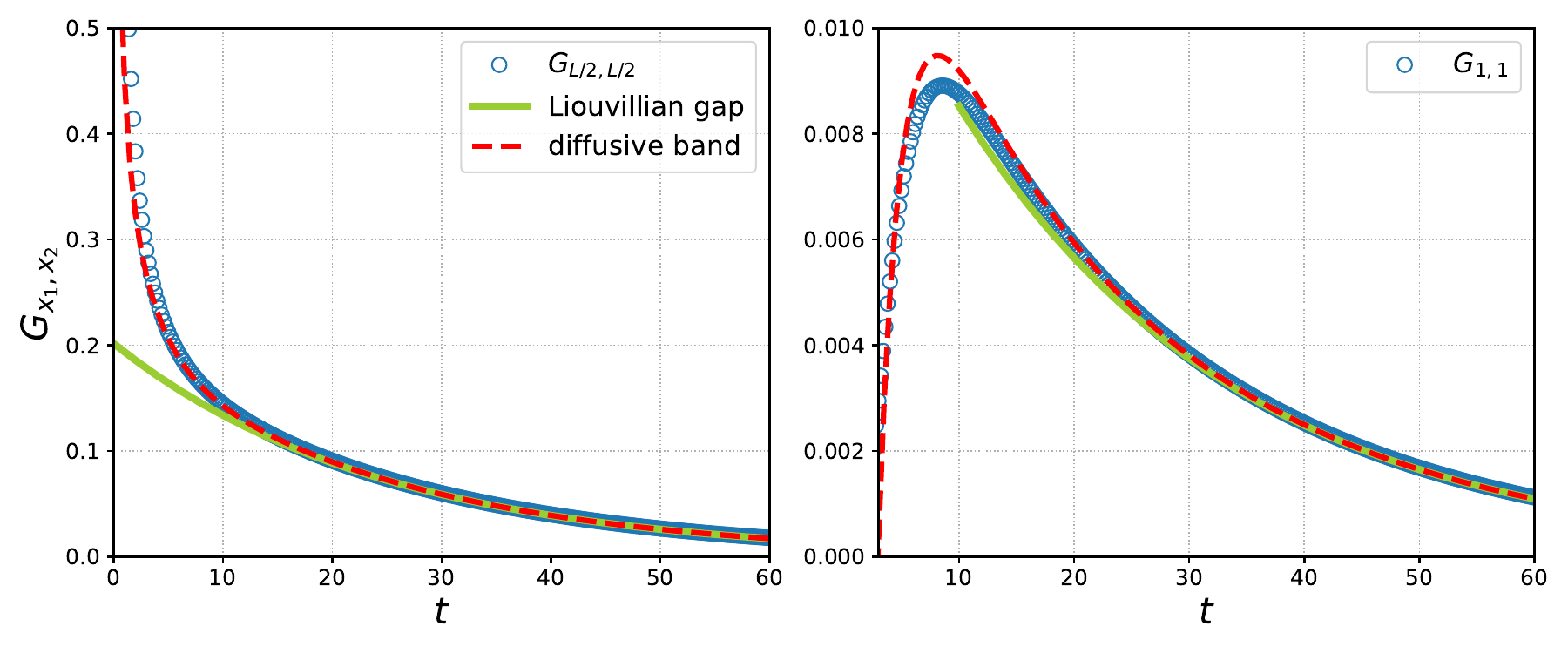}
\caption{ Evolution of the fermionic correlator $G_{x_1,x_2}$ as a function 
 of time $t$. We show results for a chain with $L=10$ sites, $\gamma=5$  and 
 $\gamma^-=1$. The correlator at $t=0$ is $G_{x_1,x_2}=\delta_{x_1,L/2}\delta_{x_2,L/2}$, 
 i.e., a fermion localized at the center of the chain. The left and right panels show 
 $G_{L/2,L/2}$ and $G_{1,1}$, respectively. The circles are exact diagonalization data. 
 The dashed line is the Bethe ansatz result obtained by solving numerically the Bethe 
 equations~\eqref{eq:beq-1} and~\eqref{eq:beq-2}, and using~\eqref{eq:ansatz}. 
 In evolving $G_{x_1,x_2}$ we only considered the eigenvalues in the diffusive 
 band (see Fig.~\ref{fig1:cartoon}), which explains the deviations from the 
 ED data. The continuous line is the result considering only the energy with the largest 
 nonzero real part. 
}
\label{fig:comparison}
\end{figure} 
%

%
\begin{figure}[t]
\centering
\includegraphics[width=.7\linewidth]{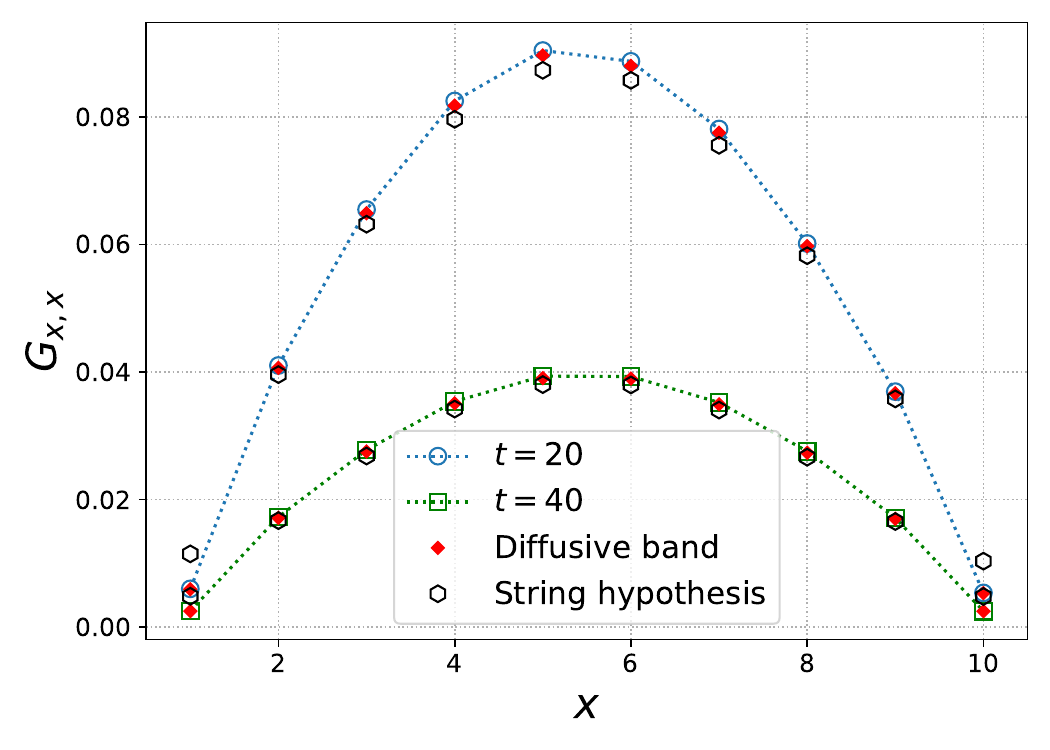}
\caption{ Density profile $G_{x,x}$ plotted as a function of the position 
 $x$ is the chain. The results are for a chain with $L=10$,  
 dissipation rate $\gamma=5$ and boundary loss rate $\gamma^-=1$. 
 The empty circles are exact diagonalization results for $t=20$ and 
 $t=40$. At $t=0$ the correlator is $G_{x_1,x_2}=\delta_{x_1,L/2}\delta_{x_2,L/2}$. 
 The empty hexagons are Bethe ansatz results in the limit $L\to\infty$. These 
 are  obtained by solving the BGT equation~\eqref{eq:bgt-eq} and using~\eqref{eq:density}. 
 The full diamonds are the same Bethe ansatz results as in  Fig.~\ref{fig:full-g}. 
}
\label{fig:profile}
\end{figure} 
%

Here we address the long-time limit of the fermionic density profile, i.e.,  the diagonal 
correlators $G_{x,x}$. This is investigated in Fig.~\ref{fig:comparison}. In the left and 
right panels we show the dynamics of $G_{L/2,L/2}$ and $G_{1,1}$, respectively. We focus on a chain with $L=10$ 
sites. The data are for $\gamma=5$ and $\gamma^-=1$. The circles in the figures are exact diagonalization 
results. The dashed line is the Bethe ansatz result obtained by using~\eqref{eq:G-exp}, where we restrict 
the sum to the eigenvalues in the diffusive band (see Fig.~\ref{fig:EDL20gp}). Importantly, we use the 
solutions of the Bethe equations~\eqref{eq:beq-1} and~\eqref{eq:beq-2}. 
Let us first focus on $G_{L/2,L/2}$. At $t=0$ one has that $G_{L/2,L2}=1$, while $G_{L/2,L/2}$ vanishes 
for $t\to\infty$. The agreement between the ED data and the Bethe ansatz is remarkable. Deviations 
are visible at short times. This is expected because of the truncation in~\eqref{eq:G-exp}. At short times 
the contribution of the complex eigenvalues of the Liouvillian cannot be neglected. The continuous line in Fig.~\ref{fig:comparison} 
is obtained by considering in~\eqref{eq:G-exp} only the contribution of the Liouvillian gap, i.e., 
the energy $\varepsilon$ with the largest nonzero real part. 
The scenario is slightly different for $G_{1,1}$ (right panel in Fig.~\ref{fig:comparison}). Precisely, 
$G_{1,1}$ is zero at $t=0$, 
it increases at later times as the particle initially at $x=L/2$ spreads towards the edges. At long 
times the dynamics is dominated by the boundary loss, and $G_{1,1}$ vanishes. 
As for $G_{L/2,L/2}$ the agreement between the ED data and the Bethe ansatz obtained by using 
the diffusive band states (dashed line) is quite satisfactory, although at intermediate times is 
only qualitatively accurate. On the other hand, the approximation obtained by restricting 
the sum in~\eqref{eq:G-exp} to the Liouvillian gap works only at long times. 

The profile of the fermionic density $G_{x,x}$ at fixed time and as a function of $x$ 
is reported in Fig.~\ref{fig:profile}. 
We show results for a chain with $L=10$ and time $t=20$ (empty circles) and $t=40$ (empty squares). 
The full diamonds are Bethe ansatz results obtained from~\eqref{eq:G-exp} restricting the sum 
over the states in the diffusive band but using the exact solutions of the Bethe equations~\eqref{eq:beq-1} 
and~\eqref{eq:beq-2}. The agreement between the Bethe ansatz and the exact diagonalization 
data is excellent for any $x$. The hexagons symbols show the Bethe ansatz results obtained employing 
the string hypothesis, i.e., by using the results of section~\ref{sec:green}. The agreement with the 
ED data is satisfactory, although some deviations are visible.

\subsection{Diffusive scaling}
\label{sec:scaling}

\begin{figure}[t]
\centering
\includegraphics[width=.95\linewidth]{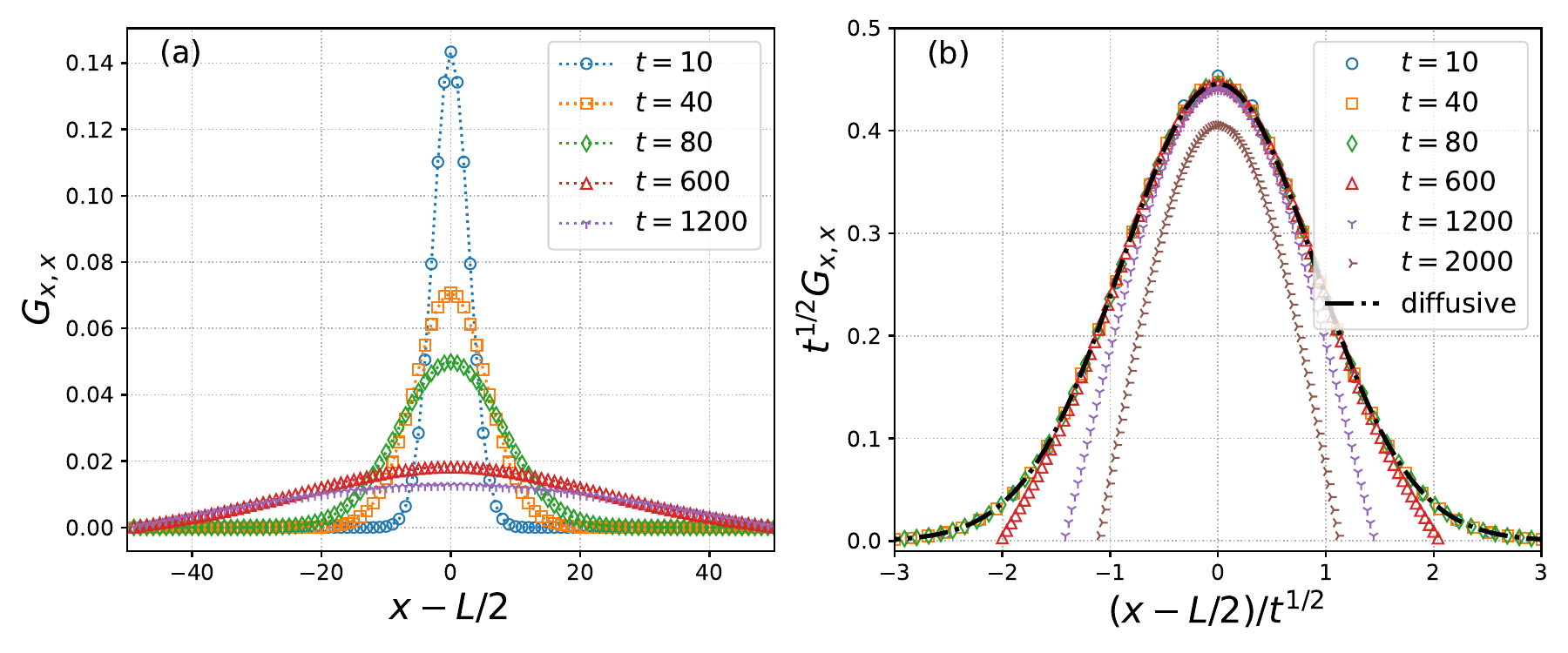}
\caption{ Dynamics of the density profile $G_{x,x}$ for a fermionic chain with $L=100$, 
 dephasing rate $\gamma=5$, and boundary loss rate $\gamma^-=1$. Panel $(a)$ shows 
 the time-dependent correlator $G_{x,x}$ as a function of $x-L/2$. At $t=0$, 
 $G_{x_1,x_2}=\delta_{x_1,L/2}\delta_{x_2,L/2}$. Panel $(b)$ shows the rescaled 
 density $t^{1/2}G_{x,x}$ as a function of $(x-L/2)/t^{1/2}$. For $t<600$ the 
 dynamics exhibits a clear diffusive scaling. At later times the boundary losss 
 affect the dynamics, and the diffusive scaling breaks down. 
}
\label{fig:scaling}
\end{figure}

In the long-time limit the density profile $G_{x,x}$ should exhibit diffusive scaling, at 
least if time is short enough that we can neglect the effect of the boundary losses. 
This diffusive behavior is observed in the periodic chain~\cite{esposito2005emergence}. 
This is investigated in Fig.~\ref{fig:scaling} focusing  on the late-time dynamics of $G_{x,x}$ 
starting from the initial condition with a fermion localized at the center of the chain. 
In Fig.~\ref{fig:scaling} $(a)$ we show $G_{x,x}$ as a function of $x-L/2$.  The symbols 
are data at different times, and are obtained by using the results of section~\ref{sec:green}. 
The data are for a system with $L=100$ sites. 
As it is clear from the Figure, the fermion density spreads diffusively as time increases, reaching the 
boundary of the chain at late times. Precisely,  in the diffusive regime $G_{x,x}$ is 
given by 
\begin{equation}
	\label{eq:scaling}
	G_{x,x}=\frac{1}{\sqrt{4\pi D t}}\exp\Big[{-\frac{(x-L/2)^2}{4Dt}}\Big],\quad D:=\frac{2}{\gamma}, 
\end{equation}
with $D$ the diffusion constant, which was derived in Ref.~\cite{esposito2005emergence}. 
The diffusive scaling is investigated in Fig.~\ref{fig:scaling} $(b)$, plotting $t^{1/2}G_{x,x}$ versus 
$(x-L/2)/t^{1/2}$. Up to  $t=80$ all the data collapse on the same curve, which is in perfect 
agreement with~\eqref{eq:scaling}. At longer times the effect of the boundary loss is non negligible 
and the diffusive scaling breaks down. At times $t\gg L^2$ the fermion density is vanishing 
at the edges of the chain, and the height of the  fermionic lump that is left around the 
center of the chain diminishes with time. 

%
%

\section{Conclusions}
\label{sec:conc}

We derived the Bethe ansatz for the spectrum of the Liouvillian ${\mathcal L}^{\scriptscriptstyle(2)}$, 
which determines the dynamics of the fermionic correlator $G_{x_1,x_2}$ in the fermionic tight-binding chain in the presence 
of bulk dephasing and boundary losses. For large enough dephasing, the spectrum of the Liouvillian comprises three different 
parts. Precisely, there are $L(L-1)/2$ complex eigenvalues that are trivially related to 
those of the tight-binding chain with boundary losses and no bulk dephasing. For this reason we dub them 
dephasing-indepedent eigenvalues. Furthermore, there are $\sim L(L-1)/2$ 
complex  eigenvalues that are perturbatively related to the dephasing-independent ones in the large chain limit. 
Finally, there is band of $\sim L$ real eigenvalues. Since the band contains the eigenvalues with the largest real parts, it 
dominates the long-time behavior of the correlator, and determines the  diffusive scaling at intermediate 
times. For this reason we dub it diffusive band. Interestingly, for large enough loss rate  boundary-related 
modes of the Liouvillian appear. Both the diffusive band and the boundary modes can be characterized by 
using the framework of the string hypothesis. Crucially, the Bethe ansatz allowed us to obtain 
the time-dependent fermionic correlator $G_{x_1,x_2}$. In particular, we provided analytic formulas 
for the long-time behavior of the fermionic density $G_{x,x}$. 

Let us now illustrate some possible directions for future work. First, we showed that despite the Liouvillian 
${\mathcal L}^{\scriptscriptstyle (2)}$ being diagonalized by Bethe ansatz, the full Liouvillian is mapped 
to the Hamiltonian of the open Hubbard chain with boundary magnetic fields and boundary pair production, which 
is not integrable. It would be interesting to investigate 
whether  the Liouvillian ${\mathcal L}^{\scriptscriptstyle(4)}$ that describes the 
dynamics of the fermionic four point function can be diagonalized by the Bethe ansatz. 
Furthermore, we showed that at long times the diffusive scaling of the fermionic 
density is broken, due to the boundary losses. It would be interesting to further investigate this regime 
to understand whether any universal scaling behavior can be extracted. An interesting direction would be to 
employ the Bethe ansatz framework to characterize the interplay between dissipation and criticality~\cite{rossini2021coherent}.  
This would require to extend the results of section~\ref{sec:asy} to non-diagonal initial conditions. Recently, 
it has been shown that several one-dimensional out-of-equilibrium systems exhibit the so-called quantum 
Mpemba effect~\cite{ares2023entanglement}. It would be interesting to investigate how the Mpemba effect  
is affected by dissipation. While this issue has been addressed numerically (see for instance Ref.~\cite{vitale2022symmetry}), 
the Bethe ansatz would allow to clarify the scenario analytically. 

The full Liouvillian describing dephasing dissipation  is not quadratic in the fermion operators. This implies that 
entanglement-related quantities are not fully determined by the two-point fermionic correlation function, in contrast with 
quadratic models~\cite{peschel2009reduced}. Still, it was observed in Ref.~\cite{alba2021spreading} 
that the ``entanglement entropies'' defined from the fermionic correlator exhibit  scaling 
behavior in the weak-dissipation hydrodynamic limit. Our results could allow to clarify the origin of this 
scaling. Finally, it would be important to understand whether the tight-binding chain with localized 
dephasing can be solved by Bethe ansatz, paving the way to characterize analytically 
entanglement scaling~\cite{alba2021unbounded,caceffo2023entanglement,chaudhari2022zeno,dabbruzzo2022logarithmic}. 

\section*{Acknowledgements}

I would like to thank Yuan Miao for useful discussions. I would also like to thank the Kavli Institute for the Physics and Mathematics 
of the Universe (Kavli IPMU), where part of this work was 
completed, for the kind hospitality. 

\appendix

\section{Derivation of the full Liouvillian}
\label{app:mapping}

Here we derive the full Liouvillian ${\mathcal L}$ governing the evolution of the density 
matrix $\rho$ (cf.~\eqref{eq:lind}). We employ the formalism of the third quantization~\cite{prosen2008third}. 
This will allow us to map the Liouvillian to a one-dimensional system of spinful fermions described by 
a Hubbard-like Hamiltonian. In section~\ref{app:hubb} we compare the result with the Hamiltonian of the one-dimensional 
Hubbard model with boundary magnetic fields. 

The action of the Liouvillian on a generic density matrix can be understood by using the formalism of 
Ref.~\cite{prosen2008third}. A generic density matrix can be decomposed as a 
superposition of strings of Majorana operators $\Gamma_{\underline\nu}$ defined as 
\begin{equation}
	\label{eq:Gamma}
	\Gamma_{\underline\nu}:= a_{1}^{\nu_1} a_2^{\nu_2}\cdots a_{2L}^{\nu_{2L}}, 
\end{equation}
where $a_j$ are Majorana fermionic operators with standard anticommutation relations $\{a_j,a_k\}=2\delta_{jk}$, 
and $\nu_j=0,1$ occupation numbers. The string of operators in~\eqref{eq:Gamma} is \emph{ordered}. 
The relationship between Majorana fermions $a_j$ and Dirac fermions $c_j$ is given as 
\begin{equation}
	a_{2j-1}=c_j+c_j^\dagger,\quad a_{2j}=i(c_j-c_j^\dagger). 
\end{equation}
It is convenient to define the creation and annihilation \emph{super operators} 
$\hat a^\dagger_j$ and $\hat a_j$ (notice  the hat) acting on $\Gamma_{\underline\nu}$ as follows 
\begin{align}
	\label{eq:hat-1}
	& \hat a^\dagger_j\Gamma_{\underline\nu}=\delta_{\nu_j,0}\pi_j\Gamma_{\underline\nu'}\\
	\label{eq:hat-2}
	& \hat a_j\Gamma_{\underline\nu}=\delta_{\nu_j,1}\pi_j\Gamma_{\underline\nu'}, 
\end{align}
where we defined the sign $\pi_j$ as 
\begin{equation}
	\pi_j:=(-1)^{\sum_{r=1}^{j-1}\nu_j}. 
\end{equation}
In~\eqref{eq:hat-1} and~\eqref{eq:hat-2}, 
we defined $\nu_r'=\nu_r$ for $r\ne j$ and $\nu'_r=1-\nu_r$ for $r=j$. The super operators $\hat a_j,\hat a_j^\dagger$ 
satisfy the standard anticommutation relations of Dirac fermions. 
First, it is straightforward to check that 
\begin{align}
	\label{eq:id-1}
	& a_j\Gamma_{\underline\nu}=(\hat a^\dagger_j+\hat a_j)\Gamma_{\underline\nu}\\
	\label{eq:id-2}
	& \Gamma_{\underline\nu}a_j=(\hat a^\dagger_j-\hat a_j)\Gamma_{\underline\nu}. 
\end{align}
In~\eqref{eq:id-2} we focus on fermionic states with  even parity, i.e., 
for which $\sum_{r=1}^{2L}\nu_r$ is even. By using~\eqref{eq:id-1} and~\eqref{eq:id-2}, 
one can easily derive the commutation relations 
\begin{align}
	\label{eq:id-3}
	& [a_j,\Gamma_{\underline\nu}]=2\hat a_j\Gamma_{\underline\nu}\\ 
	\label{eq:id-4}
	& [a_ja_k,\Gamma_{\underline\nu}]=2(\hat a_j^\dagger \hat a_k-\hat a_k^\dagger a_j)\Gamma_{\underline\nu}. 
\end{align}
For the following, it is convenient to define new fermionic super operators 
$\hat a_{\pm,j}$ as 
\begin{equation}
	\label{eq:super-c}
	\hat a_{\pm,j}:=\frac{1}{\sqrt{2}}(\hat a_{2m-1}\pm i\hat a_{2m}). 
\end{equation}
Notice that $\hat a_{\pm,j}$  and $\hat a_{+,k}$ act as ``creation'' or ``destruction'' 
super operators. For instance, $\hat a_{-,k}$ and $\hat a_{+,k}$ destroy the 
operator $ c^\dagger_k$ and $c_k$, respectively. 
Similarly, $\hat a^\dagger_{-,k}$ and $\hat a^\dagger_{+,k}$ create $c^\dagger_{k}$ 
and $c_{k}$, respectively. 

Let us decompose the Liouvillian in~\eqref{eq:lind} in a bulk contribution and in a boundary one as 
\begin{equation}
	{\mathcal L}(\Gamma_{\underline{\nu}})=
	{\mathcal L}_{\mathrm{bulk}}(\Gamma_{\underline{\nu}})
	+{\mathcal L}_{\mathrm{boundary}}(\Gamma_{\underline{\nu}}), 
\end{equation}
where ${\mathcal L}_{\mathrm{bulk}}$ contains the Hamiltonian part and the dephasing 
contribution, whereas ${\mathcal L}_{\mathrm{boundary}}$ takes into account the boundary losses. 
Specifically, the bulk Liouvillian is given as~\cite{medvedyeva2016exact,alba2021spreading} 
\begin{multline}
	\label{eq:LL-bulk}
	{\mathcal L}_{\mathrm{bulk}}=i\sum_{j=1}^{L-1}\sum_{\alpha=\pm}\alpha 
	(\hat a^\dagger_{\alpha,j}\hat a_{\alpha,j+1}+
	\hat a^\dagger_{\alpha,j+1}\hat a_{\alpha,j})\\
	+\frac{\gamma}{2}
	\sum_{j=1}^L(2\hat a^\dagger_{+,j}\hat a_{+,j}\hat a_{-,j}^\dagger\hat a_{-,j}-
	\hat a^\dagger_{+,j}\hat a_{+,j}-\hat a_{-,j}^\dagger \hat a_{-,j}), 
\end{multline}
where $\hat a_{\pm,j}$ are defined in~\eqref{eq:super-c}. 
Let us discuss the boundary term ${\mathcal L}_{\mathrm{boundary}}$. Its action on a generic 
string $\Gamma_{\underline\nu}$ reads as 
\begin{equation}
	\label{eq:l-bound}
	{\mathcal L}_{\mathrm{boundary}}(\Gamma_{\underline\nu})=\gamma^-\Big(c_1\Gamma_{\underline\nu}c_1^\dagger-
	\frac{1}{2}\{c_1^\dagger c_1,\Gamma_{\underline\nu}\}\Big)+c_1\to c_L, 
\end{equation}
where $c_j$ are Dirac fermions. 
By using~\eqref{eq:id-1} and~\eqref{eq:id-2}, Eq.~\eqref{eq:l-bound} is rewritten as 
\begin{equation}
	\label{eq:LL-1}
	{\mathcal L}_{\mathrm{boundary}}=-\frac{\gamma^-}{2}\hat a^\dagger_1\hat a_1-\frac{\gamma^-}{2}
	\hat a^\dagger_2\hat a_2
	+i\gamma^-\hat a^\dagger_1\hat a^\dagger_2+(\hat a_1,\hat a_2)\to(\hat a_{L-1},\hat a_{L}). 
\end{equation}
Finally, by using the definition of $\hat a_{+,j}$ and $\hat a_{-,j}$ in~\eqref{eq:super-c}, 
we can rewrite~\eqref{eq:LL-1} as 
\begin{equation}
	\label{eq:b-last}
	{\mathcal L}_{\mathrm{boundary}}=
	-\frac{\gamma^-}{2}\hat a^\dagger_{+,1}\hat a_{+,1}-\frac{\gamma^-}{2}\hat a^\dagger_{-,1}\hat a_{-,1}
	-\gamma^-\hat a^\dagger_{-,1}\hat a^\dagger_{+,1}+\hat a_{\pm,1}\to \hat a_{\pm,L}. 
\end{equation}
Now the first two terms in~\eqref{eq:b-last} are interpreted as boundary magnetic fields in the 
Hubbard chain. The last term, however, corresponds to creation of a pair of fermions with opposite 
spins at the boundary. 

Before proceeding, we should observe that to map ${\mathcal L}$ to a Hubbard-like Hamiltonian $H$ 
as in Ref.~\cite{medvedyeva2016exact} we have to perform a unitary transformation as 
\begin{equation}
	H=i{\mathcal U}^\dagger {\mathcal L}{\mathcal U}, 
\end{equation}
where the unitary transformation ${\mathcal U}$ is defined as~\cite{medvedyeva2016exact} 
\begin{equation}
	\label{eq:unit}
	{\mathcal U}=\prod_{\mathrm{odd}\,j}(1-2 \hat a_{-,j}^\dagger \hat a_{-,j}). 
\end{equation}
The effect of~\eqref{eq:unit} is to change the sign of the term with $\alpha=-1$ in~\eqref{eq:LL-bulk}. 
Finally, the Liouvillian is mapped to the Hubbard-like Hamiltonian $H$ as 
\begin{multline}
	\label{eq:ham-hub-1}
	H=-\sum_{j=1}^{L-1}\sum_{\sigma=\uparrow,\downarrow}(c^\dagger_{j,\sigma}c_{j+1,\sigma}+h.c.)+i\gamma
	\sum_{j}n_{j,\uparrow}n_{j,\downarrow}-i\frac{\gamma}{2}\sum_{j}(n_{j,\uparrow}+n_{j,\downarrow})\\
	-i\frac{\gamma^-}{2}(n_{1,\uparrow}+n_{1,\downarrow}+n_{L,\uparrow}+n_{L,\downarrow})
	+i\gamma^-(c_{1,\uparrow}^\dagger c_{1,\downarrow}^\dagger+(-1)^{L+1}c_{L,\uparrow}^\dagger c_{L,\downarrow}^\dagger), 
\end{multline}
where we redefined $c_{j,\uparrow}:=\hat a_{+,j}$ and $c_{j,\downarrow}:=\hat a_{-,j}$. 
Now, the Hamiltonian~\eqref{eq:ham-hub-1} is similar to that of the Hubbard chain with boundary magnetic fields. 
Precisely, Eq.~\eqref{eq:ham-hub-1} describes a chain of spinful fermions with imaginary density-density interaction, 
imaginary boundary fields. Crucially, the last term in~\eqref{eq:ham-hub-1} describes creation of pairs of fermions 
with opposite spins at the boundary of the chain. To the best of our knowledge, the boundary pair-production 
term renders the Hamiltonian~\eqref{eq:ham-hub-1} not integrable. 

However, after including a boundary fermion pump term with pump rate $\gamma^+=\gamma^-$, which is described by the Lindblad operators 
$L_{x,3}=\sqrt{\gamma^+}c^\dagger_x\delta_{x,1}$ and 
$L_{x,4}=\sqrt{\gamma^+}c^\dagger_x\delta_{x,L}$, the last term in~\eqref{eq:b-last} cancels out. 
As it was observed in Ref.~\cite{medvedyeva2016exact}, 
the resulting Hamiltonian is that of the open Hubbard chain with imaginary interactions and imaginary 
boundary magnetic fields, which is integrable~\cite{essler2005the}.

\subsection{Comparison with the Hubbard model with boundary fields}
\label{app:hubb}

Although Eq.~\eqref{eq:ham-hub-1} is not integrable in general, as we showed in section~\ref{sec:ba}, 
the Liouvillian ${\mathcal L}^{\scriptscriptstyle(2)}$ can be diagonalized by the Bethe ansatz. 
In this section we report the Bethe equations for the open Hubbard chain with boundary magnetic fields.  
We show that in the two-fermion sector the Bethe equations are the same 
as the ones derived in section~\ref{sec:ba} after some appropriate transformations. 
The fact that the pair creation terms in~\eqref{eq:ham-hub-1} do not affect the Bethe 
equations could suggest that the model is integrable by using the techniques 
of Ref.~\cite{buca2020bethe}.

The Hamiltonian of the one-dimensional Hubbard chain with open boundary conditions read as~\cite{essler2005the}  
\begin{multline}
	\label{eq:ham-hub-true}
	H=-\sum_{j=1}^{L-1}\sum_{\sigma=\uparrow,\downarrow}(c^\dagger_{j,\sigma}c_{j+1,\sigma}+h.c.) 
	+4u\sum_{j}n_{j,\uparrow}n_{j,\downarrow}-2u\sum_{j}(n_{j,\uparrow}+n_{j,\downarrow}) \\
	-p(n_{1,\uparrow}+n_{1,\downarrow})-p'(n_{L,\uparrow}+n_{L,\downarrow}), 
\end{multline}
where $c_{j,\sigma}$ are spinful fermionic operators, $n_{j,\sigma}:=c^\dagger_{j,\sigma}c_{j,\sigma}$ 
is the local fermionic density, $u$ is the interaction strength, and $p,p'$ is the strength of the boundary 
fields. Eq.~\eqref{eq:ham-hub-true} is the same as~\eqref{eq:ham-hub-1} after redefining 
$u=i\gamma/4$ and $p=i\gamma^-/2$, except for the last term in~\eqref{eq:ham-hub-1}. 
The Bethe equations for the quasimomenta $k_j$  read as~\cite{essler2005the} 
\begin{equation}
	\label{eq:bethe-hubbard}
	e^{2i k_jL}\frac{e^{ik_j}-p}{1-pe^{ik_j}}\frac{e^{ik_j}-p'}{1-p'e^{ik_j}}=
	\prod_{\ell=1}^M\frac{\sin(k_j)-\lambda_\ell+iu}{\sin(k_j)-\lambda_\ell-iu}
	\frac{\sin(k_j)+\lambda_\ell+iu}{\sin(k_j)+\lambda_\ell-iu},
\end{equation}
where $j=1,\dots,N$, together with 
\begin{equation}
	\label{eq:bethe-hubbard-1}
	\prod_{j=1}^N\frac{\lambda_\ell-\sin(k_j)+iu}{\lambda_\ell-\sin(k_j)-iu}\frac{\lambda_\ell+\sin(k_j)+iu}
	{\lambda_\ell+\sin(k_j)-iu}=\prod_{m=1,m\ne j}^M\frac{\lambda_\ell-\lambda_m+2iu}{\lambda_\ell-\lambda_m-2iu}
	\frac{\lambda_\ell+\lambda_m+2iu}{\lambda_\ell+\lambda_m-2iu}, 
\end{equation}
with $\ell=1,\dots,M$, and $N,M$ integers. The eigenvalues $\varepsilon$ of the eigenstates are given as 
\begin{equation}
	\label{eq:lambda-eq}
	\varepsilon=-\sum_{j=1}^N(2\cos(k_j)+2u). 
\end{equation}
Now, we are interested in the case with $N=2$ and $M=1$. As it is clear from the case of the 
tight-binding chain with periodic boundary conditions and bulk dephasing, the spectrum of the 
Hubbard chain with $N=2$ and $M=1$ is mapped to the spectrum of 
${\mathcal L}^{\scriptscriptstyle (2)}$. To proceed, we can solve~\eqref{eq:lambda-eq} for $\lambda_1$ to obtain 
\begin{equation}
	\label{eq:sol-lambda}
	\lambda_1=\pm\frac{1}{\sqrt{2}}(\sin^2(k_1)+\sin^2(k_2)+2u^2)^\frac{1}{2}.
\end{equation}
After substituting~\eqref{eq:sol-lambda} in~\eqref{eq:bethe-hubbard}, we obtain 
\begin{align}
	\label{eq:bethe-hubbard-1}
	& e^{2i k_1L}\frac{e^{ik_1}-p}{1-pe^{ik_1}}\frac{e^{ik_1}-p'}{1-p'e^{ik_1}}=
	\frac{\sin(k_1)-\sin(k_2)+2iu}{\sin(k_1)-\sin(k_2)-2iu}
	\frac{\sin(k_1)+\sin(k_2)+2iu}{\sin(k_1)+\sin(k_2)-2iu}\\
	\label{eq:bethe-hubbard-2}
	& e^{2i k_2L}\frac{e^{ik_2}-p}{1-pe^{ik_2}}\frac{e^{ik_2}-p'}{1-p'e^{ik_2}}=
	\frac{\sin(k_2)-\sin(k_1)+2iu}{\sin(k_2)-\sin(k_1)-2iu}
	\frac{\sin(k_2)+\sin(k_1)+2iu}{\sin(k_2)+\sin(k_1)-2iu}. 
\end{align}
After choosing $p=p'=i\gamma^-/2$ and $u=i\gamma/4$, Eq.~\eqref{eq:bethe-hubbard-1} and Eq.~\eqref{eq:bethe-hubbard-2} 
become the same as~\eqref{eq:beq-1}~\eqref{eq:beq-2} if one redefines $k_2\to k_2+\pi$. 

\bibliography{bibliography.bib}
\nolinenumbers

\end{document}